\documentclass[reprint, amsmath, amssymb, superscriptaddress]{revtex4-2}

\usepackage{times}
\usepackage{graphicx}
\usepackage{bm}
\usepackage{braket}
\usepackage{xspace}
\usepackage{float}
\usepackage{changepage} 
\usepackage{enumitem}
\usepackage{tabularx,colortbl} 
\usepackage{multirow}
\usepackage{afterpage}
\usepackage{siunitx}
\usepackage{color}
\usepackage[utf8]{inputenc}
\usepackage[T1]{fontenc}
\usepackage{comment}
\usepackage{subfiles}
\usepackage{xr}
\def \TabSampleList {S1} 
\def \TabThickness {S2} 
\def \TabRoughness {S3} 
\def \FigAllAFM {S1}
\def \FigThickness{S2} 
\def\FigRoughness{S3} 
\def \FigEDSSpectra {S4} 
\def \FigSFour {S7} 
\def \FigSFive {S8}
\def \FigPLBC {S10} 
\def \FigPLSampleF {S11} 

\begin{document}

\title{Correlated Structural and Optical Characterization of Hexagonal Boron Nitride}

\author{Jordan A. Gusdorff}
\thanks{These authors contributed equally}
\affiliation{Quantum Engineering Laboratory, Department of Electrical and Systems Engineering, University of Pennsylvania, Philadelphia, PA 19104, United States}
\affiliation{Department of Materials Science and Engineering, University of Pennsylvania, Philadelphia, PA 19104, USA}

\author{Pia Bhatia}
\thanks{These authors contributed equally}
\affiliation{Department of Physics and Astronomy, University of Pennsylvania, Philadelphia, PA 19104, USA}

\author{Trey T. Shin}
\affiliation{Department of Materials Science and Engineering, University of Pennsylvania, Philadelphia, PA 19104, USA}
\affiliation{Department of Physics and Astronomy, University of Pennsylvania, Philadelphia, PA 19104, USA}

\author{Alexandra Sofia Uy-Tioco}
\affiliation{Department of Materials Science and Engineering, University of Pennsylvania, Philadelphia, PA 19104, USA}
\affiliation{Department of Physics and Astronomy, University of Pennsylvania, Philadelphia, PA 19104, USA}

\author{Benjamin N. Sailors}
\affiliation{Department of Physics and Astronomy, University of Pennsylvania, Philadelphia, PA 19104, USA}
\affiliation{Department of Materials Science and Engineering, University of Pennsylvania, Philadelphia, PA 19104, USA}

\author{Rachael N. Keneipp}
\affiliation{Department of Physics and Astronomy, University of Pennsylvania, Philadelphia, PA 19104, USA}

\author{Marija Drndi\'c}
\affiliation{Department of Physics and Astronomy, University of Pennsylvania, Philadelphia, PA 19104, USA}

\author{Lee C. Bassett}
\affiliation{Quantum Engineering Laboratory, Department of Electrical and Systems Engineering, University of Pennsylvania, Philadelphia, PA 19104, United States}

\begin{abstract}
Hexagonal boron nitride (hBN) plays a central role in nanoelectronics and nanophotonics.
Moreover, hBN hosts room-temperature quantum emitters and optically addressable spins, making it promising for quantum sensing and quantum photonics.
Despite many investigations of their optical properties, however, the emitters’ chemical structure remains unclear, as does the role of contamination at surfaces and interfaces in forming the emitters or modifying their properties.
We prepare hBN samples that are compatible with confocal photoluminescence (PL) microscopy, transmission electron microscopy (TEM), and atomic-force microscopy (AFM), and we use those techniques to quantitatively investigate correlations between fluorescent emission, flake morphology, and surface residue.
We find that the microscopy techniques themselves induce changes in hBN's optical activity and residue morphology: PL measurements induce photobleaching, whereas TEM measurements alter surface residue and emission characteristics.
We also study the effects of common treatments\,---\,annealing and oxygen plasma cleaning\,---\,on the structure and optical activity of hBN.
The methods can be broadly applied to study two-dimensional materials, and the results illustrate the importance of correlative studies to elucidate structural factors that influence hBN's functionality as a host for quantum emitters and spin defects. 
\end{abstract}

\maketitle

\section{Introduction} \label{sec:introduction}

Hexagonal boron nitride (hBN) is a van der Waals material with broad utility.
As a wide-bandgap semiconductor ($\sim$5.9 eV) \cite{Shaik2021, Jungwirth2016}, hBN serves a key role as a dielectric layer in electronic devices \cite{Dean2010BoronElectronics,Yu2013InteractionCapacitance,Kuiri2015ProbingMeasurements}.
In addition, hBN is poised to play a key role in quantum information science, since the material hosts bright quantum emitters at room temperature \cite{Tran2016b,Exarhos2017,Mendelson2019EngineeringNitride,Patel2022}.
Some emitters display spin-dependent optical properties and can be utilized as spin qubits \cite{Exarhos2019,Chejanovsky2021,Guo2023CoherentTemperature,Patel2024RoomNitride,Stern2024AConditions}.
However, hBN's quantum emitters are not yet fully understood.
Identifying emitters in two-dimensional materials is especially complex because they are sensitive to interfaces \cite{Exarhos2017,Akbari2021} and other local environmental effects (\textit{e.g.}, defects \cite{Chejanovsky2021}, strain \cite{Proscia2018,Sajid2020}, and contamination \cite{Yang2022EffectMaterials,Li2023ProlongedEmitters}).
Understanding and controlling the properties of hBN, especially at its surface, are therefore crucial to advancing quantum information and photonics applications.

Techniques such as photoluminescence (PL) microscopy and scanning electron microscopy (SEM) are commonly employed to characterize hBN.
PL provides information about optical properties, including absorption and emission characteristics.
For the study of quantum emitters, it provides essential chemical and electronic information \textit{via} the spectra of electronic and vibronic transitions \cite{Jungwirth2016,Exarhos2017,Wigger2019}, as well as spatial information regarding the emitters' density and brightness \cite{Breitweiser2020}.
SEM has been used to study surface topography and morphology \cite{Exarhos2017,Mendelson2019EngineeringNitride}, in addition to activating emission centers in hBN \cite{Tran2016b,Breitweiser2020,Fournier2021Position-controlledNitride}.
More recently, transmission electron microscopy (TEM) has been used as a means of probing quantum emission in hBN, particularly when paired with PL or cathodoluminescence spectroscopy \cite{Hayee2020,Li2023ProlongedEmitters}.
TEMs enable direct observation of material structure down to the atomic level and are generally operated at much higher accelerating voltages than SEMs ($\sim$80--200 kV versus $\sim$5--30 kV), opening up a new regime of irradiation conditions to be explored for the possible creation of emitters \cite{Alem2009AtomicallyMicroscopy,Bui2023CreationIrradiation,Keneipp2024NanoscaleBeam}.

While the optical properties and material structure of hBN are often studied separately, few studies have investigated them in parallel.
The absence of studies employing both TEM and PL microscopy can be partially attributed to the disparity in scale at which these two techniques operate.
Confocal PL studies in the visible-wavelength range are typically limited by diffraction to a resolution of $\sim$300 nm, whereas TEM studies can achieve atomic resolution.
By combining both of these techniques, the capabilities of TEM can be leveraged to elucidate the structural and chemical composition of quantum emitters identified via PL microscopy.

\begin{figure*}
    \centering
    \includegraphics[width=1\linewidth]{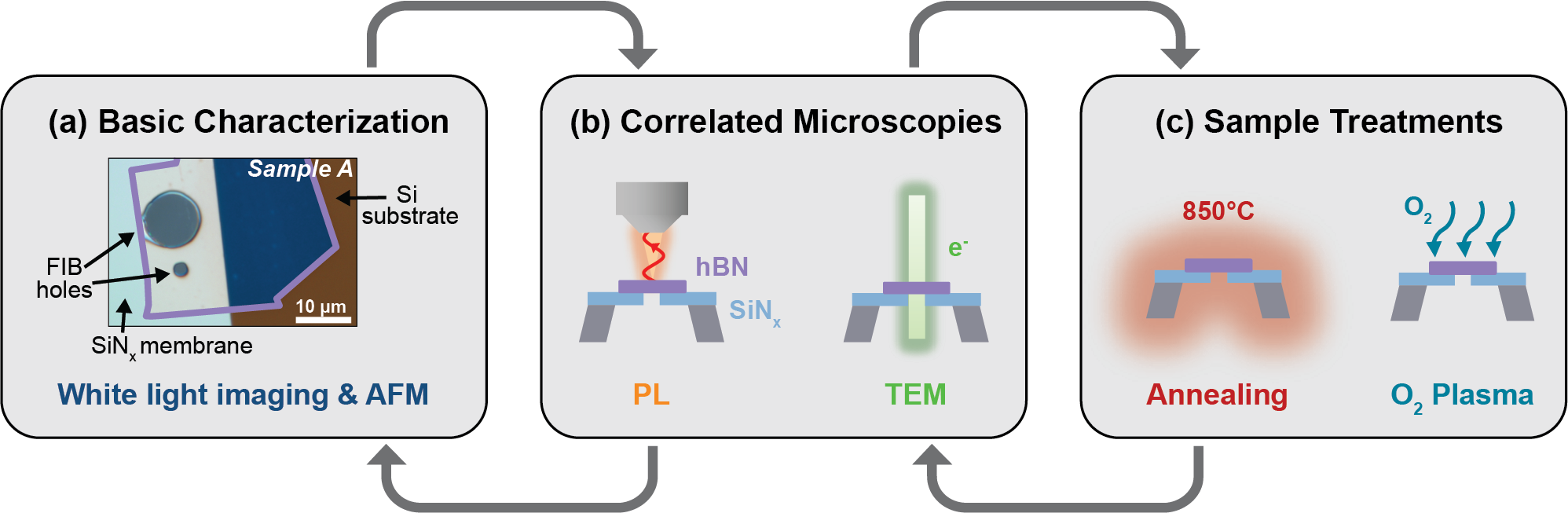}
    \caption{
    \textbf{Summary of Experimental Methods}
    HBN samples undergo an iterative process that includes basic characterization, correlative PL and TEM measurements, and treatments. (a) All samples are first characterized with white light imaging and AFM. The white light image shows Sample A, a mechanically exfoliated flake of hBN (outlined in purple) suspended over a 10 $\mu$m diameter hole milled in a silicon nitride membrane.
    (b) Basic characterization is followed by PL and/or TEM measurements. The schematics illustrate a cross-sectional view of each correlated microscopy technique.
    (c) Some samples then receive annealing and/or oxygen plasma treatments. Changes in optical activity and morphology following treatment are analyzed via the correlated microscopies and basic characterization methods.}
    \label{fig:Schematic}
\end{figure*}

In this work, we combine PL and TEM imaging together with atomic-force microscopy (AFM) to bridge the gap in understanding between hBN's optical properties and material structure.
We study changes in optical activity and flake morphology associated with different measurements and treatments, focusing on overall statistical properties of the material rather than investigating individual single-photon emitters with PL or few-atom defects with TEM.
To this end, we fabricate custom substrates compatible with both microscopies onto which hBN flakes are transferred with thicknesses ranging from $\sim$10--30 nm.
First, we consider the effects of PL and TEM measurements themselves.
Using quantitative methods, we show that exposure to the electron beam during TEM imaging induces changes to surface contaminants and that PL imaging induces photobleaching.
Subsequently, we study correlated changes in structure and optical activity following annealing and oxygen plasma, treatments commonly applied to hBN \cite{Li2019PurificationTreatments,Li2023ProlongedEmitters,Venturi2024SelectiveNitride,Shaik2021,Mohajerani2024NarrowbandNitride, Chen2023AnnealingNitride}.
Both treatments significantly alter sample morphology and optical activity.
We discuss the implications of this work for understanding and engineering quantum emitters in hBN, as well as for future investigations of other two-dimensional materials.

\section{Material and Instrumental Considerations} \label{sec:considerations}

\subsection{Sample Requirements}
Figure \ref{fig:Schematic} summarizes the sample geometry along with the different characterization methods and treatments we explore in this work.
We report results for six hBN samples, labeled A through F, each of which received a distinct sequence of measurements and treatments that are summarized in Table~\TabSampleList\ in the Supporting Information.
Sample preparation begins by mechanically exfoliating flakes of hBN from bulk crystal (HQ Graphene) and using polydimethylsiloxane (PDMS) viscoelastic stamping to transfer flakes onto 100$\times$100 $\mu$m$^2$ silicon nitride (SiN$_x$) windows.
To improve TEM resolution and ensure that PL measurements are not subject to substrate induced effects, hBN flakes are suspended over holes patterned in SiN$_x$ windows via focused ion beam (FIB) milling.
A 5- to 10-$\mu$m-diameter FIB hole (Fig.~\ref{fig:Schematic}a) provides a sufficiently large region of suspended hBN suitable for both TEM and PL microscopy.
Smaller, 2-$\mu$m-diameter FIB holes (Fig.~\ref{fig:Schematic}a) help define the sample orientation and provide an area over which the TEM can initially be aligned.

Flake thickness is another important consideration.
Monolayer or few-layer flakes are best for TEM studies at atomic resolution.
However, stable single-photon emitters in hBN tend to occur in thicker flakes that are typically at least 30 nm thick \cite{Tran2016,Ngoc2018EffectsNitride}.
Therefore, we utilize flakes of intermediate thickness, ranging from $\sim$10 nm to $\sim$30 nm, to meet the needs of both microscopies.
Flakes of the desired thickness are identified by optical contrast prior to transferring and subsequently confirmed with AFM (see Tables \TabThickness--\TabRoughness\ and Figs.~\FigAllAFM--\FigRoughness\ 
 in the Supporting Information).

\subsection{Effects of PL and TEM imaging}

\begin{figure*}
    \centering
    \includegraphics[width=1\linewidth]{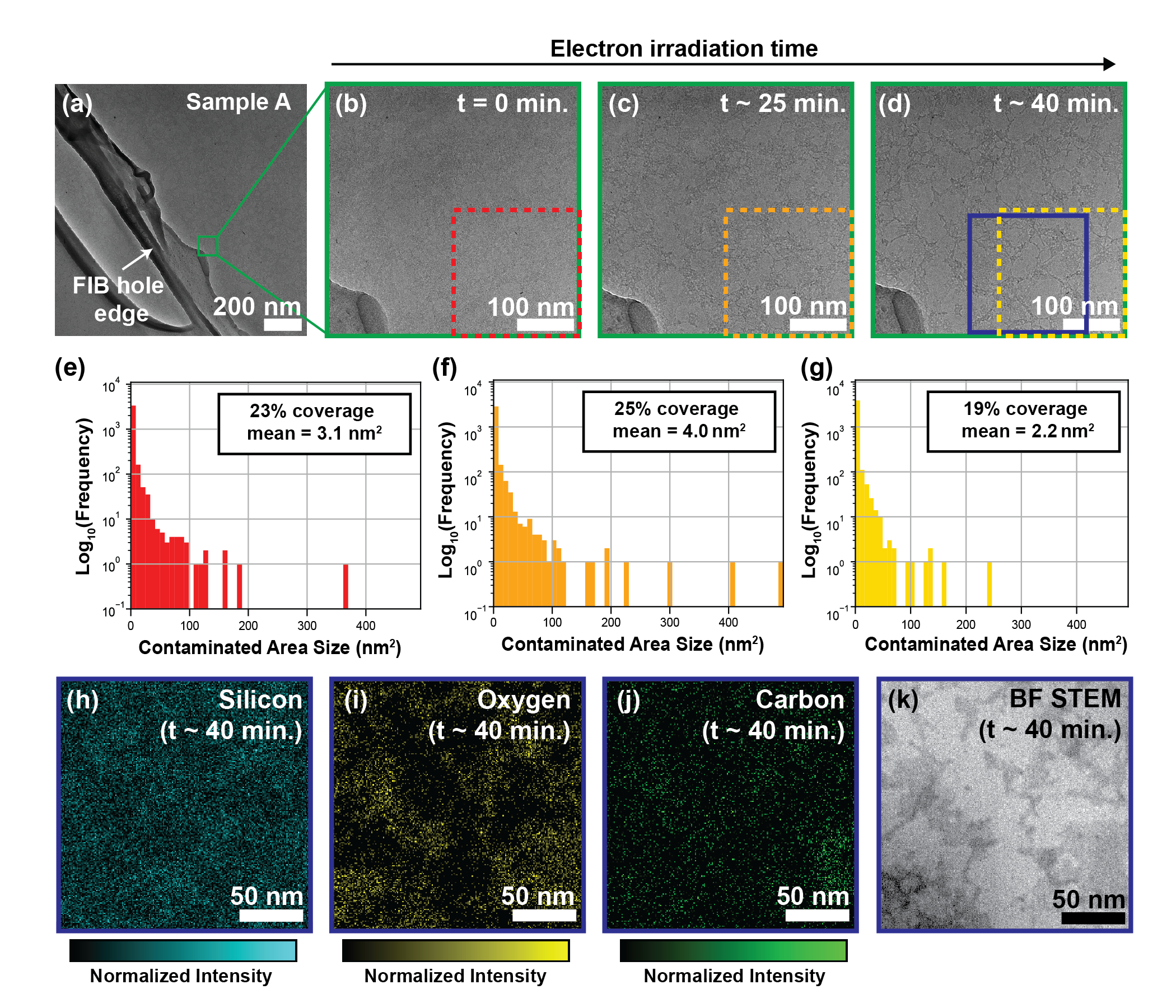}
    \caption{
    \textbf{Effect of TEM Electron Irradiation on hBN Contaminants}
    (a) Low magnification TEM micrograph of Sample A. The green box denotes the region shown in b-d. (b) Initial TEM micrograph of a suspended region of hBN. The region boxed in red was analyzed to produce the histogram in panel (e). (c) TEM micrograph of the same region after $\sim$25 minutes of exposure to the electron beam. The region boxed in orange was analyzed to produce the histogram in panel (f). (d) TEM micrograph of the same region after $\sim$40 minutes of exposure to the electron beam, illustrating a change in surface contamination. The region boxed in yellow was analyzed to produce the histogram in panel (g). The blue box corresponds to panels e-g. (h-j) EDS elemental maps taken of Sample A. (k) Scanning TEM (STEM) brightfield micrograph of the same region shown in the elemental maps.
    }
    \label{fig:fig2}
\end{figure*}

Figure \ref{fig:fig2} demonstrates the effect of electron beam irradiation in the TEM on a representative region of Sample A over time.
The sample is initially covered with a film of residue, as seen in Figure \ref{fig:fig2}b.
However, with prolonged exposure to the electron beam ($\sim$20 nA probe current), the residue changes in structure and density.
Qualitatively, the residue appears to clump together and diminish over time, resulting in the lacey network of residue shown in Figure \ref{fig:fig2}d after 40 minutes of irradiation.
We quantitatively analyze the residue using an image-processing algorithm to detect regions of contamination and characterize its spatial distribution, as described in Sec.~IV of the Supporting Information. 
Figures~\ref{fig:fig2}e-g show the contamination area distributions corresponding to each image along with the overall surface coverage.
Interestingly, we observe that both the surface coverage and mean area increase slightly over the first 25 minutes of TEM exposure, and then decrease after 40 minutes.
The overall decrease in contamination coverage and average size between 0 and 40 minutes of irradiation shows that prolonged exposure to the electron beam has a cleaning effect, while the increase between 0 and 25 minutes implies that this effect is dose-dependent.
It has been shown that hydrocarbons tend to diffuse across the surface of samples towards the electron beam where they are eventually polymerized and fixed in place, often covering features of interest \cite{Egerton2004RadiationSEM,Rykaczewski2007AnalysisMicroscopy}.
On the other hand, electron irradiation can also remove material through knock-on damage, radiolysis, or charging effects \cite{Egerton2019RadiationTEM}.
These competing phenomena can explain the non-monotonic changes we observe.

The residue shown in Figure \ref{fig:fig2}b-d likely arises from the PDMS polymer used to transfer hBN flakes onto the substrates.
A wide body of literature describes how contamination arising from polymer-assisted transfer techniques of 2D materials is a pervasive issue for the field \cite{Bhatia2024ANitride, Cheliotis2024AMaterials, Jain2018MinimizingPDMS, Jang2023AGraphene}.
Energy dispersive x-ray spectroscopy (EDS) maps confirm the presence of silicon, carbon and oxygen species, which is consistent with the chemical composition of PDMS (see Fig.~\ref{fig:fig2}h-j and Fig.~\FigEDSSpectra\ in the Supporting Information).
Moreover, all three maps appear to be spatially correlated with the contaminated (dark) areas in the corresponding brightfield scanning TEM (BF STEM) image (Fig.~\ref{fig:fig2}k).
In recent related work, we used electron energy loss spectroscopy (EELS) and aberration corrected STEM to study the composition and spatial distribution of surface contamination that results from transfer of hBN using PDMS and poly bis-A carbonate (PC) polymers  \cite{Bhatia2024ANitride}.
We found that both transfer methods yield flakes with significant and comparable residue, underscoring the importance of further research to improve transfer methods and to establish reliable cleaning protocols.

\begin{figure*}
    \centering
    \includegraphics[width=1\linewidth]{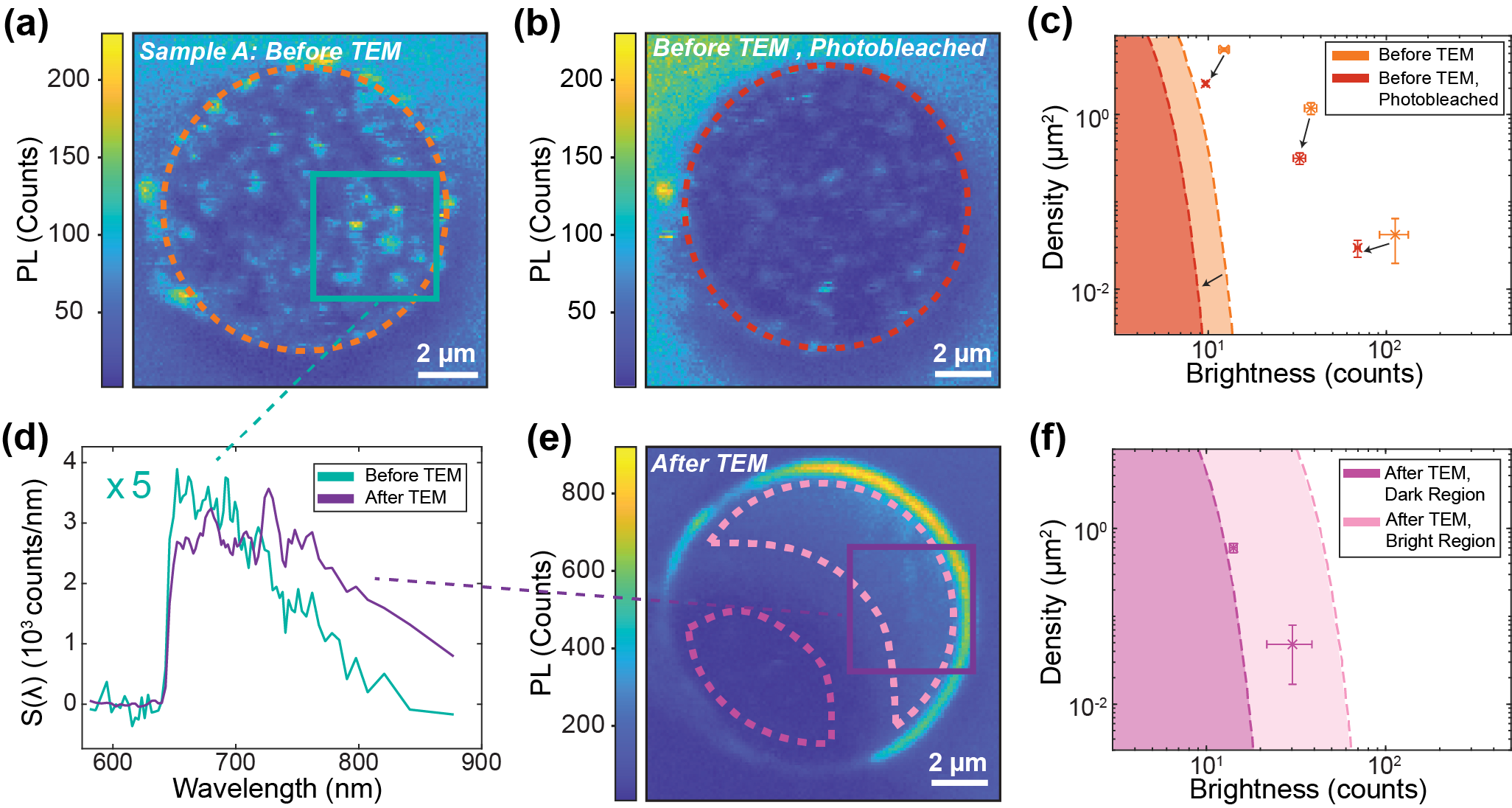}
    \caption{
    \textbf{Measurement Effects on Optical Properties}
    (a) Initial PL image of Sample A. Orange dashed circle outlines suspended area analyzed in (c). Teal square outlines area of PL spectrum collection.
    (b) PL image after photobleaching. Red dashed circle outlines area analyzed in (c).
    (c) Emitter family analysis of Sample A before observations in the TEM. Arrows indicate the likely evolution of the emitter families due to photobleaching.
    (d) Spatial PL spectra corresponding to the initial PL image (teal curve) and the PL image after TEM measurements (purple curve).
    (e) PL image after TEM measurements. Light pink and dark pink dashed outlines distinguish the bright and dark regions, respectively, which are analyzed in panel (f).
    (f) Emitter family analysis after TEM measurements.
    }
    \label{fig:PL-effects}
\end{figure*}

To understand how TEM electron irradiation and associated changes to surface contaminants impact optical activity, Sample A was also studied with confocal PL microscopy before and after TEM imaging.
Figure \ref{fig:PL-effects} summarizes the results.
The first confocal PL image prior to electron irradiation shows bright localized emission (Fig.~\ref{fig:PL-effects}a).
The 592-nm laser was then rastered over the 12$\times$12 $\mu$m$^2$ area at 100 $\mu$W for about 20 minutes, and then at 300 $\mu$W for about 2 minutes.
The subsequent PL image (Fig.~\ref{fig:PL-effects}b) shows that after prolonged laser exposure, emission is still localized but is substantially dimmer within the suspended region.
This phenomenon, known as photobleaching, is ubiquitous to quantum emitters and has been previously observed in hBN \cite{Li2023ProlongedEmitters}.
Figure~\ref{fig:PL-effects}c shows the results of a statistical analysis of these data that identifies families of emitters based on their density and brightness (see \citet{Breitweiser2020} and Methods for details).
The results indicate a decrease in emitter density or brightness by a factor of $\sim$3 following the laser exposure, accompanied by a decrease in the overall PL background.
In order to asses the spectral properties of the emitter ensembles, optical spectra were acquired while continuously rastering the confocal spot over the $\sim5\times5$~$\mu$m rectangular region marked in Fig.~\ref{fig:PL-effects}a.
The resulting spatially-averaged spectrum (teal curve in Fig.~\ref{fig:PL-effects}d) is consistent with the typical distribution of visible quantum emitters in these samples \cite{Tran2016b,Exarhos2017,Patel2022}.
 
After photobleaching, Sample A was subjected to electron irradiation in the TEM to obtain the data shown in Figure~\ref{fig:fig2}b-d (see Table \TabSampleList\ for the  full sequence of measurements), and then returned to the optical microscope.
Figure~\ref{fig:fig2}e illustrates the dramatic qualitative change in the fluorescence properties following TEM imaging.
The overall fluorescent intensity increases by a factor of $\sim$4, and a bright ring appears at the edges of the suspended region.
The upper right portion of the suspended region exhibits diffuse bright emission, whereas the lower left portion---where Sample A was irradiated for $\sim$40 minutes---appears significantly dimmer.
The brightening and shift from spatially localized to diffuse emission is consistent across other TEM-imaged samples in this study (Fig.~\FigPLBC b,e), as well as findings in prior work \cite{Keneipp2024NanoscaleBeam}.
The subsequent dimming with extended electron irradiation, however, is a notable exception to the trend.
The non-monotonic change in optical activity may be related to the non-monotonic changes in PDMS residue observed in the TEM images of Sample A.
Previous authors have even proposed organic compounds as the source of visible emission in hBN \cite{Neumann2023OrganicMica}.
It is also notable that the 80kV accelerating voltage used here exceeds the knock-on energies of boron and nitrogen \cite{Keneipp2024NanoscaleBeam, Kotakoski2010ElectronMonolayers, Dai2023EvolutionNitride, Gilbert2017FabricationNitride}.
Therefore, the optical variations could reflect changes in chemical bonding of quantum emitters, or variations in hBN's Fermi level that affect the emitter charge states.

Emitter family analysis of the PL image taken after TEM measurements reveals the presence of two emitter groups in the darker region (Fig.~\ref{fig:PL-effects}f).
These groups may have evolved from those groups that were initially present before TEM irradiation (Fig.~\ref{fig:PL-effects}c), however direct identification is unclear.
The overall trend is a decrease in both the density and brightness of emitters, whereas the brightness level of uniform background PL has increased.
Since the emitter family analysis assumes a random spatial distribution, it cannot be applied to the bright region of Figure \ref{fig:PL-effects}e, because it exhibits a clear spatial gradient in brightness.
The background level of this bright region, which may actually reflect a dense ensemble of emitters, is indicated in Figure \ref{fig:PL-effects}f for reference.
Meanwhile, the spatially averaged emission spectrum shows a shift towards longer wavelengths following electron irradiation (purple curve in Fig.~\ref{fig:PL-effects}d).
This is further evidence in support of changes to the nature of emitters due to TEM exposure, whether that be direct, \textit{e.g.}, through the creation of new emitters, or indirect, \textit{e.g.}, through modulation of the material's Fermi level or predominant optical relaxation pathways.

\section{Treatment Effects} 

To investigate the effects of sample treatments on the hBN, we study separate samples in parallel using TEM and PL microscopy to decouple the effects of TEM on PL emission. 
We perform two treatments (oxygen plasma and annealing) on hBN and study the resulting optical and structural changes.
Oxygen plasma and annealing are commonly used to improve quantum emission in hBN (\textit{e.g.}, increasing brightness and density of emitters) \cite{Breitweiser2020,Tran2016b,Fischer2021}, but their effects on the morphology of the material are not well understood.
In this section, we study four samples of hBN (Samples B through E). 
These samples are comparable in thickness (see Table \TabThickness) and were fabricated using identical protocols. 

\subsection{TEM Analysis of Treatments}
\begin{figure*}
    \centering
    \includegraphics[width=1\linewidth]{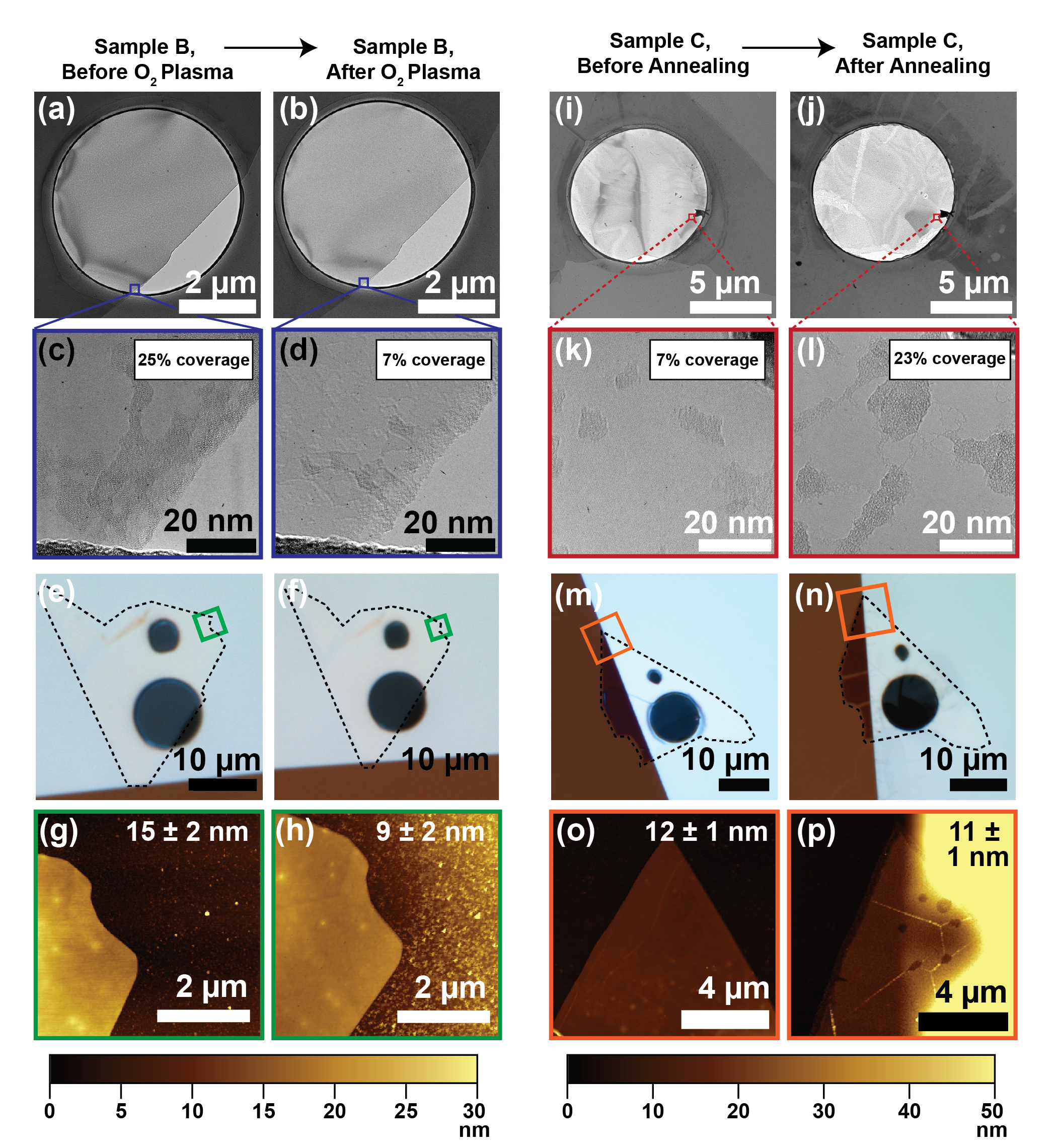}
    \caption{
    \textbf{Effects of annealing on Sample B and O$_2$ plasma on Sample C, characterized by TEM and AFM}
    (a-d) TEM images of Sample B before and after oxygen plasma treatment. The navy boxes in (a) and (b) correspond to (c) and (d), respectively.
    (e-f) White light images before and after oxygen plasma treatment. The hBN flake is outlined in black, and the green boxes indicate the AFM region.
    (g-h) AFM scans of Sample B before and after oxygen plasma treatment.
    (i-l) TEM images of Sample C before and after annealing. The red boxes in (i) and (j) correspond to (k) and (l), respectively. 
    (m-n) White light images before and after annealing. The hBN flake is outlined in black, and the orange boxes indicate the AFM region.
    (o-p) AFM scans of Sample C before and after annealing.}
    \label{fig:fig4}
\end{figure*}

Figure \ref{fig:fig4} summarizes morphological changes to the exfoliated hBN flakes after oxygen plasma and annealing.
Sample B was treated with oxygen plasma for 5 minutes (50 W, 50 sccm O$_2$).
TEM images show that on the micron scale, this treatment smoothed wrinkles on the flake (Fig.~\ref{fig:fig4}a,b) but had no other noticeable effects.
On the nanometer scale, however, oxygen plasma significantly reduced contamination coverage on Sample B from 25\% to 7\%, with a similar decrease in average contaminant area size in the representative region analyzed (Fig.~\ref{fig:fig4}c,d and \FigSFour). 
Furthermore, TEM images indicate that etching of hBN occurred, evidenced by the appearance of triangular-shaped defects following oxygen plasma treatment (Fig.~\ref{fig:fig4}d).
This treatment also resulted in significant thinning of Sample B (15 $\pm$ 2 nm $\rightarrow$ 9 $\pm$ 2 nm) and a decrease in surface roughness by a factor of $\sim$2 (see Tables \TabThickness--\TabRoughness\ and Figs.~\FigThickness--\FigRoughness).
These changes are consistent with the removal of surface contaminants along with etching of hBN. 
The former result was expected, given that oxygen plasma treatment is regularly employed as a cleaning method \cite{Na2021ModulationTreatment, Jadwiszczak2021PlasmaApplications, Kim2016EffectsMoS2}, however ideally the treatment should not etch the hBN.
The Supporting Information includes further data on Samples E and F, which were treated with oxygen plasma using a lower oxygen flow rate and longer time, indicating that the amount of etching can be reduced using these parameters.

In comparison, Sample C was annealed for 2 hours at 850~\textdegree C in an argon environment.
TEM images show that annealing resulted in significant cracking and warping of the flake (Fig.~\ref{fig:fig4}i,j).
Annealing also resulted in the formation of sub-micron holes, visible in the bottom right corner of Fig.~\ref{fig:fig4}j and in the AFM data of Fig.~\ref{fig:fig4}p. 
These changes may result from differing rates of thermal expansion between the hBN flake and the underlying SiN$_x$ membrane \cite{Zhang2023TheAnnealing}.
Further analysis of TEM images reveal a local increase in contamination coverage (from 7\% to 23\%) and average contaminant size (from 18 nm$^2$ to 56 nm$^2$) on Sample C (see Figs.~\ref{fig:fig4}k,l and \FigSFive). 
This implies that the annealing treatment either augments contamination or, at minimum, induces a substantial redistribution of contaminants on the hBN surface. 
These changes are unexpected since annealing, like oxygen plasma, is often utilized as a cleaning technique \cite{Garcia2012EffectiveDevices, Lin2012GrapheneBe}.
AFM measurements of Sample C show that the annealing treatment results in a $\sim$2.5-fold increase in roughness without significant thinning (Fig.~\ref{fig:fig4}o,p and Tables \TabThickness--\TabRoughness).

\subsection{PL Analysis of Treatments}

\begin{figure*}
    \centering
    \includegraphics[width=1\linewidth]{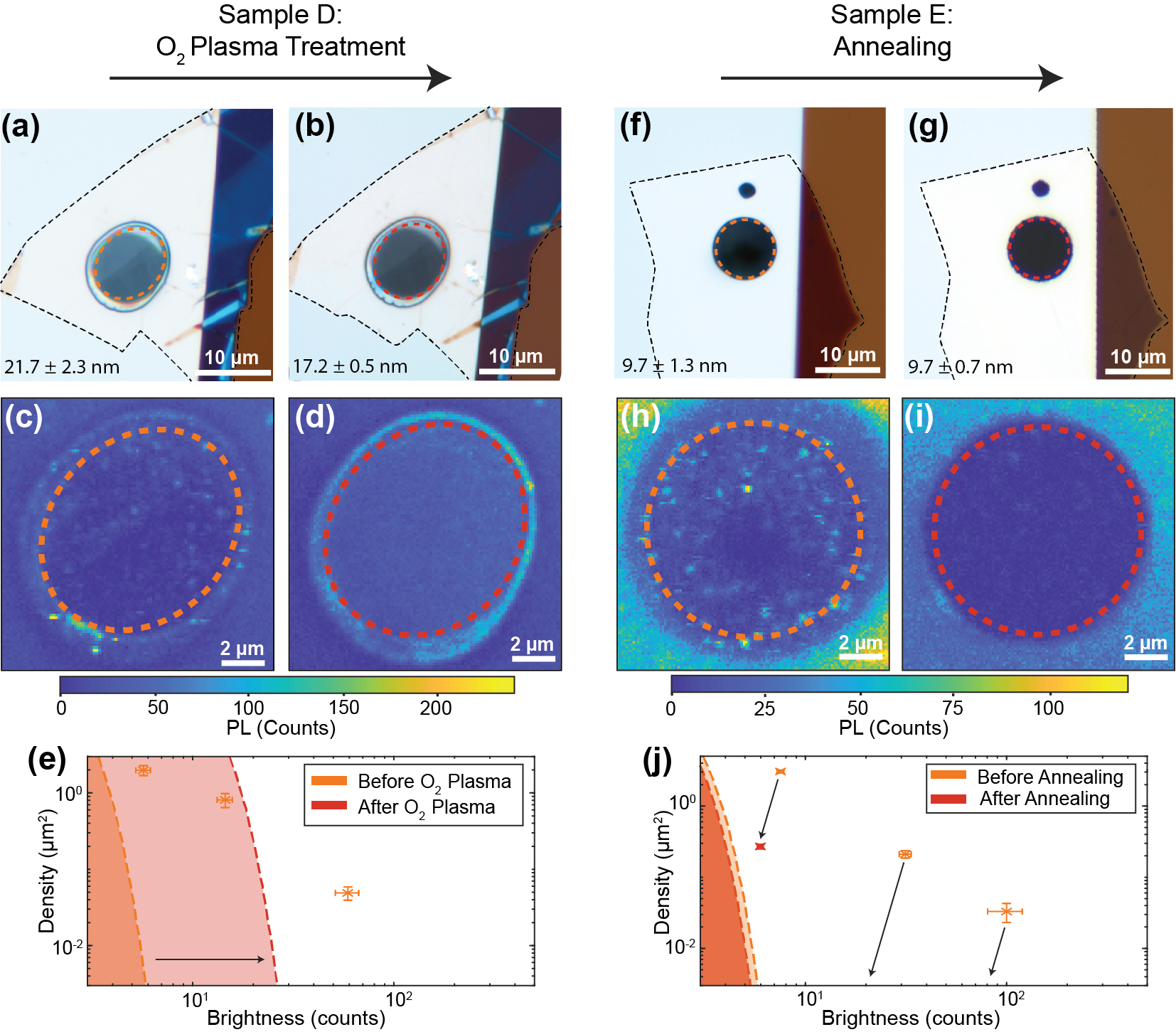}
    \caption{
    \textbf{Effects of annealing and O$_2$ plasma on optical measurements}
    (a-b) White light images showing Sample D (hBN flake outlined in black) before (a) and after (b) oxygen plasma treatment. Orange and red dashed ovals outline the area suspended over the FIB hole.
    (c-d) PL images of Sample D before (c) and after (d) oxygen plasma treatment. Areas within orange and red dashed ovals are analyzed in (e).
    (e) Emitter family analysis of Sample D before and after oxygen plasma treatment. Arrow indicates the increase in background emission due to the treatment.
    (f-g) White light images showing Sample E (hBN flake outlined in black) before (f) and after (g) annealing. Orange and red dashed circles outline the area suspended over the FIB hole.
    (h-i) PL images of Sample E before (h) and after (i) annealing. Areas within orange and red dashed circles are analyzed in (j).
    (j) Emitter family analysis of Sample E before and after annealing. Arrows indicate a proposed evolution of the emitter families.
    }
    \label{fig:PL-treatments}
\end{figure*}

Figure \ref{fig:PL-treatments} shows how comparable annealing and oxygen plasma treatments affect hBN's optical activity.
Like Sample B, Sample D was exposed to oxygen plasma for 5 minutes (50 W, 50 sccm O$_2$).
While there are no obvious changes visible in the white-light optical image, AFM reveals that the flake becomes thinner (Fig.~\ref{fig:PL-treatments}a,b).
The initial PL image shows localized emission over the suspended region (Fig.~\ref{fig:PL-treatments}c), similar to other samples.
After the oxygen plasma treatment, the PL becomes bright and diffuse without any localized emission visible (Fig.~\ref{fig:PL-treatments}d).
Quantitative analysis uncovers three emitter families initially present in the sample (Fig.~\ref{fig:PL-treatments}e).
After oxygen plasma treatment, a spatial gradient prevents analysis of the emitter families.
However, we observe that the background level increases significantly.

The brightening and shift to a diffuse morphology observed in Sample D were unexpected since oxygen plasma treatments are typically used as a cleaning procedure \cite{Griffiths2010QuantificationCleaning,Hugenschmidt2022Electron-Beam-InducedMitigation,Mitchell2015ContaminationMicroscopy}.
However, these changes could be attributed to damage from etching or oxygen implantation.
AFM measurements revealed that the sample is thinned (Table \TabThickness), and the treatment is the same as for Sample B, for which obvious evidence of etching was observed in TEM (Fig.~\ref{fig:fig4}a-d).
Potentially, damage from oxygen plasma treatment could increase the background brightness similarly to electron irradiation in the TEM.
The effects differ, however, in that oxygen plasma primarily affects the material’s surface, whereas electron irradiation can impact the bulk \cite{Vogl2019AtomicNitride}.
The majority of reports only consider oxygen plasma treatment followed by annealing, as it has been suggested that both treatments are necessary for stable emitter formation \cite{Fischer2021}.
Another possible explanation for the increase in brightness is that oxygen ions are implanted during the irradiation process, changing hBN's chemical composition.
Previous studies have shown that oxygen ion implantation can alter the optical properties of hBN, including a shift to a diffuse morphology \cite{Mendelson2021}.

The Supporting Information includes further examples of changes in PL following oxygen plasma treatment of other samples (Fig.~\FigPLBC--\FigPLSampleF).
The samples consistently exhibited a shift to diffuse emission, however the changes in brightness were less consistent, indicating that the treatment does not affect all samples equally.
Sample D exhibited the greatest increase in brightness.
When using a lower oxygen flow rate for a longer time, we observed negligible changes to the PL characteristics  (Fig.~\FigPLSampleF) and less thinning in AFM (Fig.~\FigThickness).
These observations highlight the future potential to tune the treatment recipe as required to achieve a desired outcome.

Sample E was annealed at 850~\textdegree C for two hours in an argon environment, using the same procedure as for Sample C.
After the treatment, the sample maintains its thickness and appears largely unchanged under white light, although slight variations in color that are apparent before annealing appear to diminish afterwards (Fig.~\ref{fig:PL-treatments}f,g).
AFM measurements show that the surface roughness of Sample E increased substantially upon annealing, consistent with the results of an identical annealing treatment applied to Sample C.
Interestingly, Sample E was subjected to an oxygen plasma treatment (50 W, 25 sccm O$_2$, 10 min) following annealing, and yet the surface roughness remains more than twice as high as the pre-treatment measurement.
The initial PL image (Fig.~\ref{fig:PL-treatments}h) displays the same background pattern as its corresponding white light image (Fig.~\ref{fig:PL-treatments}f), superimposed with localized emitters.
After the sample is annealed, however, few localized emitters remain, and they appear almost uniform with the background (Fig.~\ref{fig:PL-treatments}i).
Emitter family analysis identifies three distinct groups before Sample E was annealed and only one group afterwards, with negligible changes to the background (Fig.~\ref{fig:PL-treatments}j).
The one remaining emitter group is dimmer than all of the groups from before the treatment.
This result contrasts with previous studies, as annealing is regularly used to stabilize and brighten emitters in hBN \cite{Breitweiser2020,Tran2016b}.
In this case, emission becomes dimmer, and the few localized emitters that remain are unstable.
These qualitative features are consistent across all samples observed in this study.

The thinness of the samples may help to explain the discrepancy, since prior works on quantum emitters in exfoliated hBN generally consider much thicker flakes ($\gtrsim100$~nm).
Selection bias may play a role as well, since annealing's brightening effect might apply mainly to crystals that already host stable emitters, and samples without improved optical properties are excluded from such reports.
If quantum emission arises from lattice defects, the decrease in PL intensity could correspond to annealing repairing the lattice. 
Even though the TEM image of Sample C shows global warping of the sample (Fig.~\ref{fig:fig4}i,j), annealing would still be expected to heal the lattice on the local atomic scale.
The suspended region of Sample C has no discernible PL signal after annealing, confirming that warping does not by itself induce brightening (Fig.~\FigPLBC f).
Another explanation for the dimming effect is defect migration or transmutation, which potentially reduces the density of emitters or blue-shifts their excitation or emission wavelength to become undetectable in our experimental setup \cite{Venturi2024SelectiveNitride,Weston2018NativeNitride}.
 
\section{Conclusions and Outlook} \label{sec:conclusions}
In this work, we demonstrated a method for sequential, correlated TEM and PL measurements of multi-layer hBN, and we used these complementary techniques, along with AFM and white light imaging, to study the interplay between surface contamination, overall morphology, flake thickness, and optical properties of mechanically exfoliated hBN\,---\,all critical considerations towards scalable quantum information and photonics technologies utilizing hBN.
We show that viscoelastic stamping with PDMS results in considerable residue on the surface of hBN flakes through direct TEM imaging and elemental mapping.
We observe that the residue exhibits non-monotonic changes in structure, and the optical emission exhibits non-monotonic changes in brightness as a result of electron irradiation.
In these samples, all optical emission is unstable and subject to photobleaching, potentially due to the flake thickness.
We find that oxygen plasma treatment results in thinning of hBN and reduces surface residue while brightening and diffusing PL emission.
In contrast, annealing induces warping and cracking in hBN flakes and results in an increase or re-distribution of surface residue while reducing optical activity overall.
These results contradict previous assumptions that such treatments clean the surface by removing residue without inflicting damage on the hBN, highlighting the need for further investigations of this subject.

Such work could include a systematic comparison of different sample preparation methods (\textit{i.e.}, dry transfer, wet transfer, \textit{etc.}) and investigations of the resulting residue or lack thereof; a systematic comparison of the effect of thickness on optical activity, especially in combination with thermal annealing; fine-tuning of oxygen plasma parameters to optimally clean surfaces without damaging the hBN matrix; and investigation of large-scale morphological (\textit{e.g.}, strain-induced) effects by utilizing different sample substrates and geometries and annealing recipes.
Motivated by the broad and interdisciplinary use of hBN, such studies can significantly advance the realization of new nanoelectronic and nanophotonic technologies. 
Likewise for use in quantum information science and photonics, quantitative understanding and control of surface contamination is essential to engineering materials and devices. 
By correlating optical and structural imaging, it will become possible to characterize and potentially even create quantum emitters at the atomic scale, facilitating the design and realization of new generations of quantum technologies based on hBN. 
Moreover, the methods and procedures  introduced in this work can be directly adapted to study the structural and optical properties of a broad range of 2D materials.

\section*{Methods} \label{sec:methods}

\subsection{Sample Preparation}
TEM grids were purchased from SiMPore with 50-nm-thick SiN$_x$ membranes suspended above silicon supports.
The suspended SiN$_x$ region was 100 $\times$ 100 $\mu$m large.
Holes of varying sizes (1, 5, and 10 $\mu$m diameter) were drilled in the SiN membranes with a Xe$^+$ ion beam in a Tescan S8252X plasma focused ion beam scanning electron microscope (PFIB-SEM).
The PFIB was operated at 15 kV and 300 pA for drilling and the SEM at 5 kV and 100 pA for imaging of SiN memebranes.
Once patterned with holes of the desired size and location, we mechanically exfoliated hBN (HQ Graphene) and transferred multi-layer flakes onto the SiN membranes with PDMS stamping.
This procedure has been previously reported; see Bhatia \textit{et al.}\cite{Bhatia2024ANitride} and Keneipp \textit{et al.}\cite{Keneipp2024NanoscaleBeam} for more details.

\subsection{Atomic Force Microscopy}
AFM measurements were performed using a Bruker Icon AFM.
To obtain values for flake thickness and roughness, regions of the flake comparable in thickness (as determined by white-light imaging) to the suspended region were scanned.
In general, AFM scans improved when taken areas of the flake supported by the silicon substrate, instead of regions of the flake on the SiN$_x$ membrane, as the thin SiN$_x$ membrane resulted in a strong, nonuniform background signal in the AFM scans.
Additional information about AFM data processing are provided in the Supporting Information.

\subsection{Optical Characterization}
PL images were collected using a custom-built confocal microscope with a 592 nm continuous-wave laser (MPB Communications, VF-P-200- 592).
The sample was illuminated with circularly polarized light at 100 $\mu$W unless otherwise noted.
A long-pass filter with a cut-on wavelength of 650 nm (Semrock, BLP01- 635R-25) is placed in the collection path to eliminate scattered light.
The collected emission goes through a multimode fiber that is either routed to single-photon counting modules (Laser Components, Count T-100) or a spectrometer (Princeton Instruments, Iso-Plane160 and Pixis 100 CCD).
For further details, see Patel \textit{et al.}\cite{Patel2022}.

\subsection{Emitter Family Analysis}
The analysis used to categorize emitter groups is based on previous work by Breitweiser \textit{et al.}\cite{Breitweiser2020}.
Briefly, we consider a histogram of the pixel intensity for selected regions of PL images (as indicated by dashed ovals in Figures \ref{fig:PL-effects} and \ref{fig:PL-treatments}).
Fits of the histogram determine the background level and the properties of each emitter group.
We select the number of emitter groups based on the reduced chi-squared value of the overall fit, and we extract density and brightness parameters from the emitter groups.

\subsection{TEM Characterization}
TEM characterization was performed in the JEOL JEM-F200 S/TEM equipped with a Gatan OneView Camera and JEOL dual SDD detectors for EDS spectra acquisition. The JEOL F200 was operated at 80kV in both TEM mode and STEM mode. All TEM data were processed using Gatan DigitalMicrograph and ImageJ.

\subsection{TEM Residue Analysis}
TEM micrographs for Samples A, B and C were analyzed using an image processing routine to identify regions of residue on each flake. 
This image processing routine is described in detail in Sections IV and V of the Supporting Information. 
Once contaminated regions were identified, histograms of contaminated area size were plotted (see Figs. \ref{fig:fig2}e-g) to quantify the changes observed in TEM micrographs. 

\subsection{Sample Treatments}
Oxygen plasma treatment was performed on hBN samples in an Anatech SCE 108 Barrel Asher. TEM grids with hBN flakes were secured to a 4-inch, 500 $\mu$m Si and 1500 \r{A} silicon oxide wafer using kapton tape. This wafer was placed into the process chamber with oxygen plasma at 50W with flow rate of 25 or 50 sccm for 5 or 10 minutes, respectively.

Annealing treatment was performed on hBN samples in the Carbolite HZS 1200\textdegree C 3 Zone Split Tube Furnace.
The tube used is dedicated to hBN use only and cleaned prior to treatment with isopropyl alcohol and pressurized nitrogen gas.
Samples are placed in an alumina ceramic boat and annealed in an Ar environment at 850\textdegree C for two hours.

\section{Acknowledgements} \label{sec:acknowledgements}
J.A.G. and L.C.B. acknowledge support from the NSF under awards DMR-1922278 and DMR-2019444.
J.A.G. also acknowledges support from an NSF Graduate Research Fellowship (DGE-1845298).
P.B., T.T.S., A.S.U., B.N.S., R.N.K., and M.D. acknowledge support from DOE grant DE-SC0023224 on advanced in-situ TEM analysis and ex-situ studies of 2D hBN. P.B. was also funded by NIH grant R01HG012413 for the development of hBN/graphene nanopores.
This work was carried out in part at the Singh Center for Nanotechnology, which is supported by the NSF National Nanotechnology Coordinated Infrastructure Program under grant NNCI-2025608.
The authors also acknowledge use of facilities and instrumentation supported by NSF through the University of Pennsylvania Materials Research Science and Engineering Center (MRSEC) (DMR-1720530).
Finally, the authors also wish to thank Dr. Douglas Yates and Dr. Jamie Ford for their assistance with TEM characterization and FIB milling of substrates.

\bibliography{references}

\end{document}


\renewcommand{\figurename}{Figure}
\renewcommand{\thefigure}{S\arabic{figure}}
\renewcommand{\tablename}{Table}
\renewcommand{\thetable}{S\arabic{table}}

\def \MainFigTwo {2}
\def \MainFigFour {4} 
\def \MainFigPlEffects {3}
\def \MainFigPLTreatments {5}

\title{Supporting Information for Correlated Structural and Optical Characterization of Hexagonal Boron Nitride}

\author{Jordan A. Gusdorff}
\thanks{These authors contributed equally}
\affiliation{Quantum Engineering Laboratory, Department of Electrical and Systems Engineering, University of Pennsylvania, Philadelphia, PA 19104, United States}
\affiliation{Department of Materials Science and Engineering, University of Pennsylvania, Philadelphia, PA 19104, USA}

\author{Pia Bhatia}
\thanks{These authors contributed equally}
\affiliation{Department of Physics and Astronomy, University of Pennsylvania, Philadelphia, PA 19104, USA}

\author{Trey T. Shin}
\affiliation{Department of Materials Science and Engineering, University of Pennsylvania, Philadelphia, PA 19104, USA}
\affiliation{Department of Physics and Astronomy, University of Pennsylvania, Philadelphia, PA 19104, USA}

\author{Alexandra Sofia Uy-Tioco}
\affiliation{Department of Materials Science and Engineering, University of Pennsylvania, Philadelphia, PA 19104, USA}
\affiliation{Department of Physics and Astronomy, University of Pennsylvania, Philadelphia, PA 19104, USA}

\author{Benjamin N. Sailors}
\affiliation{Department of Materials Science and Engineering, University of Pennsylvania, Philadelphia, PA 19104, USA}
\affiliation{Department of Physics and Astronomy, University of Pennsylvania, Philadelphia, PA 19104, USA}

\author{Rachael N. Keneipp}
\affiliation{Department of Physics and Astronomy, University of Pennsylvania, Philadelphia, PA 19104, USA}

\author{Marija Drndi\'c}
\affiliation{Department of Physics and Astronomy, University of Pennsylvania, Philadelphia, PA 19104, USA}

\author{Lee C. Bassett}
\affiliation{Quantum Engineering Laboratory, Department of Electrical and Systems Engineering, University of Pennsylvania, Philadelphia, PA 19104, United States}
\maketitle

\section{Summary of Sample Treatments and Measurements}
\label{sec:samplesummary}
\begin{table}[hbt!] 
    \centering
    \setlength{\tabcolsep}{5pt}
    \renewcommand{\arraystretch}{2}
    \begin{tabular}{|c|c|c|} \hline 
        Sample & Figure & Characterization and Treatment Sequence \\\hline 
        \vspace{-9pt}
        \multirow{2}*{A}& Fig.~1, 2, 3, & \multirow{2}*{AFM $\rightarrow$ PL $\rightarrow$ TEM$^\dagger$ $\rightarrow$ PL $\rightarrow$ TEM $\rightarrow$ PL $\rightarrow$ TEM$^{\dagger \dagger}$}\\
        & \ref{fig:S1}, \ref{fig:S2}, \ref{fig:S3}, \ref{fig:TEM-vacuum}& \\\hline

        B& Fig.~4, \ref{fig:contaminationB}, \ref{fig:PL-BC}& AFM $\rightarrow$ PL $\rightarrow$ TEM $\rightarrow$ PL $\rightarrow$ O$_2$ Plasma (5m) $\rightarrow$ PL $\rightarrow$ TEM $\rightarrow$ PL\\\hline

        C& Fig.~4, \ref{ContaminationC}, \ref{fig:PL-BC}& AFM $\rightarrow$ PL $\rightarrow$ TEM $\rightarrow$ PL $\rightarrow$ Anneal $\rightarrow$ PL $\rightarrow$ TEM $\rightarrow$ PL $\rightarrow$ AFM\\\hline 

        D& Fig.~5 & AFM $\rightarrow$ PL $\rightarrow$ O$_2$ Plasma (5m) $\rightarrow$ AFM $\rightarrow$ PL $\rightarrow$ Anneal $\rightarrow$ PL\\\hline

        E& Fig.~5 & AFM $\rightarrow$ PL $\rightarrow$ Anneal $\rightarrow$ PL $\rightarrow$ O$_2$ Plasma (10m) $\rightarrow$ AFM $\rightarrow$ PL \\ \hline 

        F& Fig.~\ref{fig:alt-oxygen} & AFM $\rightarrow$ PL $\rightarrow$ O$_2$ Plasma (10m) $\rightarrow$ PL $\rightarrow$ Anneal $\rightarrow$ PL $\rightarrow$ AFM\\ \hline

    \end{tabular}
    \caption{\textbf{Summary of Sequential Measurements and Treatments}}
    \label{tab:sample-list}
\end{table}

$^\dagger$ See Figure \ref{fig:TEM-vacuum} for further details. 
$^{\dagger \dagger}$ EDS measurements taken. 

\section{Summary of Sample Thickness and Roughness} 
\label{sec:AFM}

Each sample was measured using atomic force microscopy (AFM). All AFM scans are included in Fig.~\ref{fig:SXAllAFM} below. 
AFM was performed using a Bruker Dimension Icon AFM, and data was processed using Gywddion.
In order to obtain a height with error, four height profiles were measured on each flake, each with 50 data points and approximately 1 $\mu$m long.
The data points corresponding to the substrate were averaged, and the data points corresponding to the flake were averaged. The difference of these values is the reported flake height. 
The sum of the standard deviations in substrate height and flake height is reported as the error. 
The initial and final thicknesses of all samples are summarized in Table \ref{tab:thickness} and Fig.~\ref{fig:ThicknessPlot}. 

Similarly, surface roughness values for Samples B-E were obtained using the Statistical Quantities Tool in Gwyddion. 
Their initial and final RMS roughnesses ($Sq$) and arithmetic mean roughnesses ($Sa$) are listed in Table \ref{tab:roughness}. 
These results are also summarized visually in Fig.~\ref{fig:RoughnessPlot}
These values were obtained by selecting the same four regions on each flake before and after treatment. Roughness measurements were obtained from each region and averaged, with the standard deviation reported as the error.
Suitable regions were those that appeared featureless (no cracks, wrinkles, etc.). 

\newpage
\begin{figure}[htb!]
    \centering
    \includegraphics[width=0.75\linewidth]{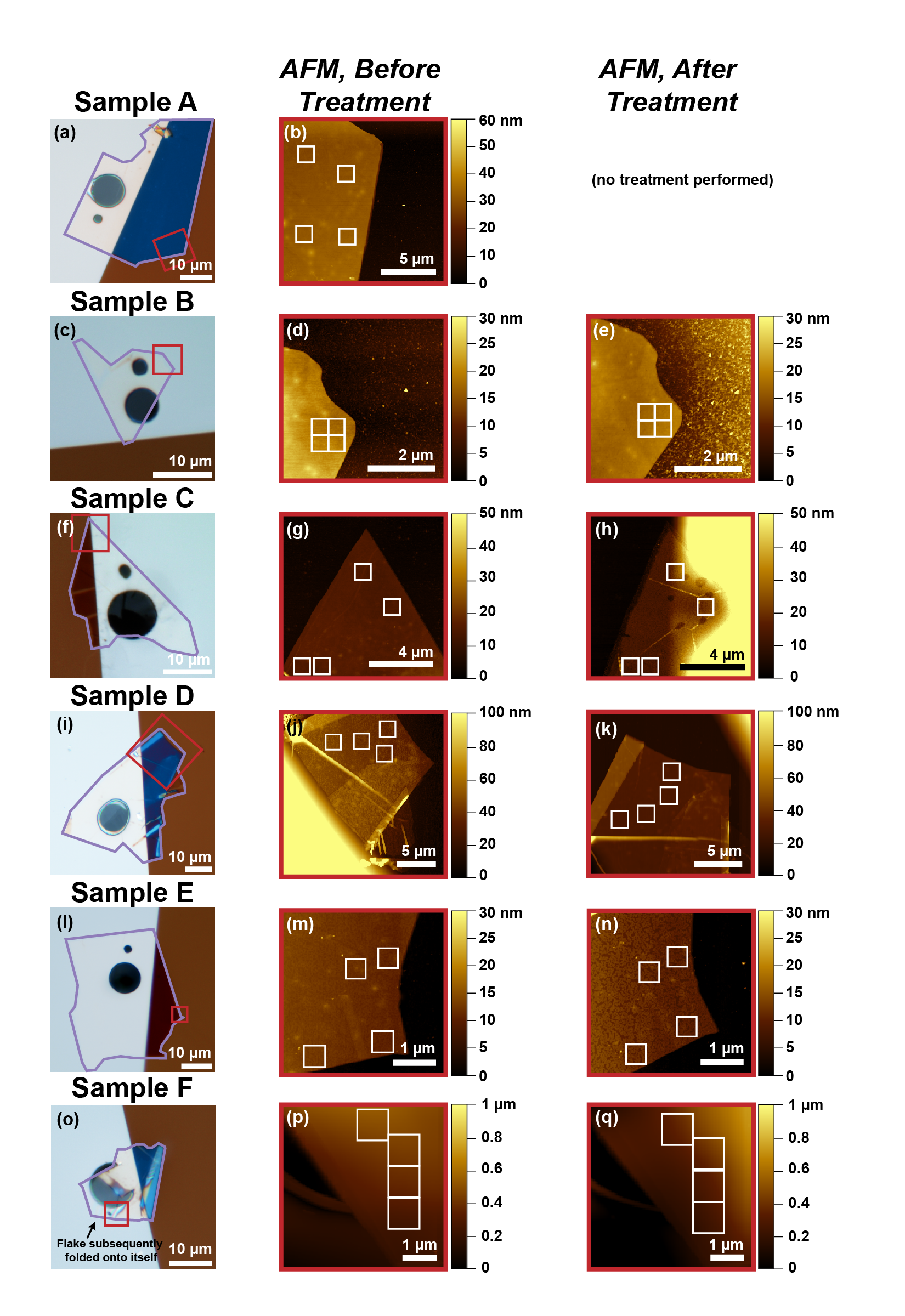}
    \caption{This figure shows the AFM scans, before and after treatment, for all samples. HBN flakes are outlined in purple. Red boxes on the white-light images correspond to the regions measured with AFM. White boxes on AFM scans correspond to the areas used to determine the roughness of each sample.}
    \label{fig:SXAllAFM}
\end{figure}

\begin{table}[htb!]
    \centering
    \setlength{\tabcolsep}{5pt}
    \renewcommand{\arraystretch}{2}
    \begin{tabular}{|c|c|c|c|c|} \hline 
    \vspace{-15pt}
         \multirow{2}*{Sample} & Initial & \multirow{2}*{Treatment} &Final &Change in\\
         & Thickness (nm) & & Thickness (nm) & Thickness (nm)\\\hline 
         
         A& 27.7 $\pm$ 0.8 & --&-- &--\\\hline 
         
         B& 15.3 $\pm$ 1.6 & O$_2$ Plasma (5m)&9.2 $\pm$ 1.9 &6.2 $\pm$ 3.5\\\hline
         
         C& 11.7 $\pm$ 0.7 & Anneal &11.3 $\pm$ 1.1 &0.4 $\pm$ 1.8\\\hline 

         D& 21.7 $\pm$ 2.3 & O$_2$ Plasma (5m)&17.2 $\pm$ 0.5 &4.5 $\pm$ 2.8\\\hline
         
         E& 9.7 $\pm$ 1.3 & Anneal, then O$_2$ Plasma (10m)&9.7 $\pm$ 0.7 &0.0 $\pm$ 2.0\\\hline
        
        F& 17.9 $\pm$ 0.8 & O$_2$ Plasma (10m), then Anneal&14.6 $\pm$ 1.1 &3.3 $\pm$ 1.9\\\hline 
        
    \end{tabular}
    \caption{\textbf{Summary of Sample Thickness}}
    \label{tab:thickness}
\end{table}
\begin{figure}
    \centering
    \includegraphics[width=0.75\linewidth]{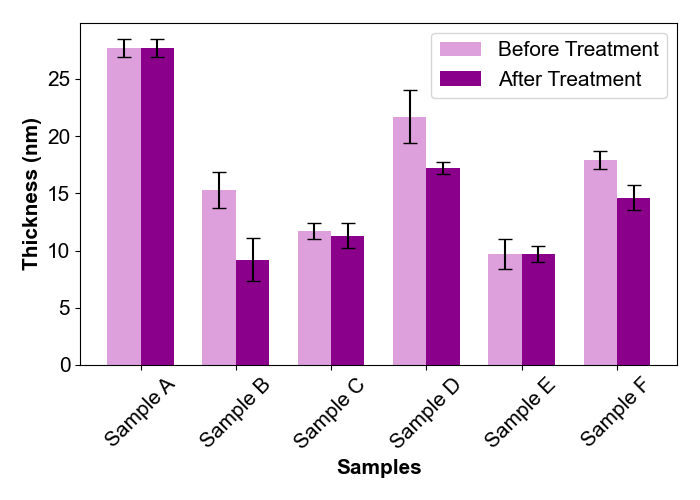}
    \caption{\textbf{Summary of Sample Thicknesses: }A plot of sample thickness, before (light pink) and after treatment (dark pink). }
    \label{fig:ThicknessPlot}
\end{figure}
\def\mathbi#1{\textbf{\em #1}} 
\clearpage
\begin{table}[hbt!]
    \centering
    \setlength{\tabcolsep}{5pt}
    \renewcommand{\arraystretch}{2}
    \begin{tabular}{|c|c|c|c|c|} \hline
    
    \vspace{-9pt}
    \multirow{2}*{Sample} & Pre Treatment $Sq$ (pm)  &\multirow{2}*{Treatment}& Post Treatment $Sq$ (pm) & $Sq$ Change Factor\\
     & \textbf{Pre Treatment $\mathbi{Sa}$ (pm)}  && \textbf{Post Treatment $\mathbi{Sa}$ (pm)} & \textbf{$\mathbi{Sa}$ Change Factor}\\
    \hline

    \vspace{-9pt}
    \multirow{2}*{A} & 490 $\pm$ 23  &\multirow{2}*{-}
& \multirow{2}*{-} & \multirow{2}*{-}\\
     & \textbf{385 $\pm$ 20}  &&&\\
    \hline

    \vspace{-9pt}
    \multirow{2}*{B} & 701 $\pm$ 64  &\multirow{2}*{O$_2$ Plasma (5m)}
& 460 $\pm$ 22 & 0.66 $\pm$ 0.07\\
     & \textbf{563 $\pm$ 56}  && \textbf{359 $\pm$ 18} & \textbf{0.64 $\pm$ 0.07}\\
    \hline

    \vspace{-9pt}
    \multirow{2}*{C} & 434 $\pm$ 9  &\multirow{2}*{Anneal} 
& 1145 $\pm$ 71 & 2.64 $\pm$ 0.17\\
     & \textbf{348 $\pm$ 3}  && \textbf{906 $\pm$ 54} & \textbf{2.60 $\pm$ 0.16}\\
    \hline

    \vspace{-9pt}
    \multirow{2}*{D} & 2900 $\pm$ 220  &\multirow{2}*{O$_2$ Plasma (5m)}
& 1240 $\pm$ 270 & 0.43 $\pm$ 0.10\\
    & \textbf{2370 $\pm$ 180}  && \textbf{1020 $\pm$ 240} & \textbf{0.43 $\pm$ 0.11}\\
    \hline

    \vspace{-9pt}
    \multirow{2}*{E} & 460 $\pm$ 36  &Anneal $\rightarrow$& 994 $\pm$ 36 & 2.16 $\pm$ 0.18\\
    & \textbf{361 $\pm$ 25}  &O$_2$ Plasma (10m)& \textbf{807 $\pm$ 33} & \textbf{2.24 $\pm$ 0.18}\\
    \hline

    \vspace{-9pt}
    \multirow{2}*{F} & 8950 $\pm$ 690  &O$_2$ Plasma (10m) $\rightarrow$& 8650 $\pm$ 160 & 0.97 $\pm$ 0.08\\
    & \textbf{7590 $\pm$ 520}  &Anneal& \textbf{6960 $\pm$ 180} & \textbf{0.92 $\pm$ 0.07}\\
    \hline
    
    \end{tabular}
    \caption{\textbf{Summary of Sample Roughness}}
    \label{tab:roughness}
\end{table}
\begin{figure}[htb!]
\centering
    \includegraphics[width=1\linewidth]{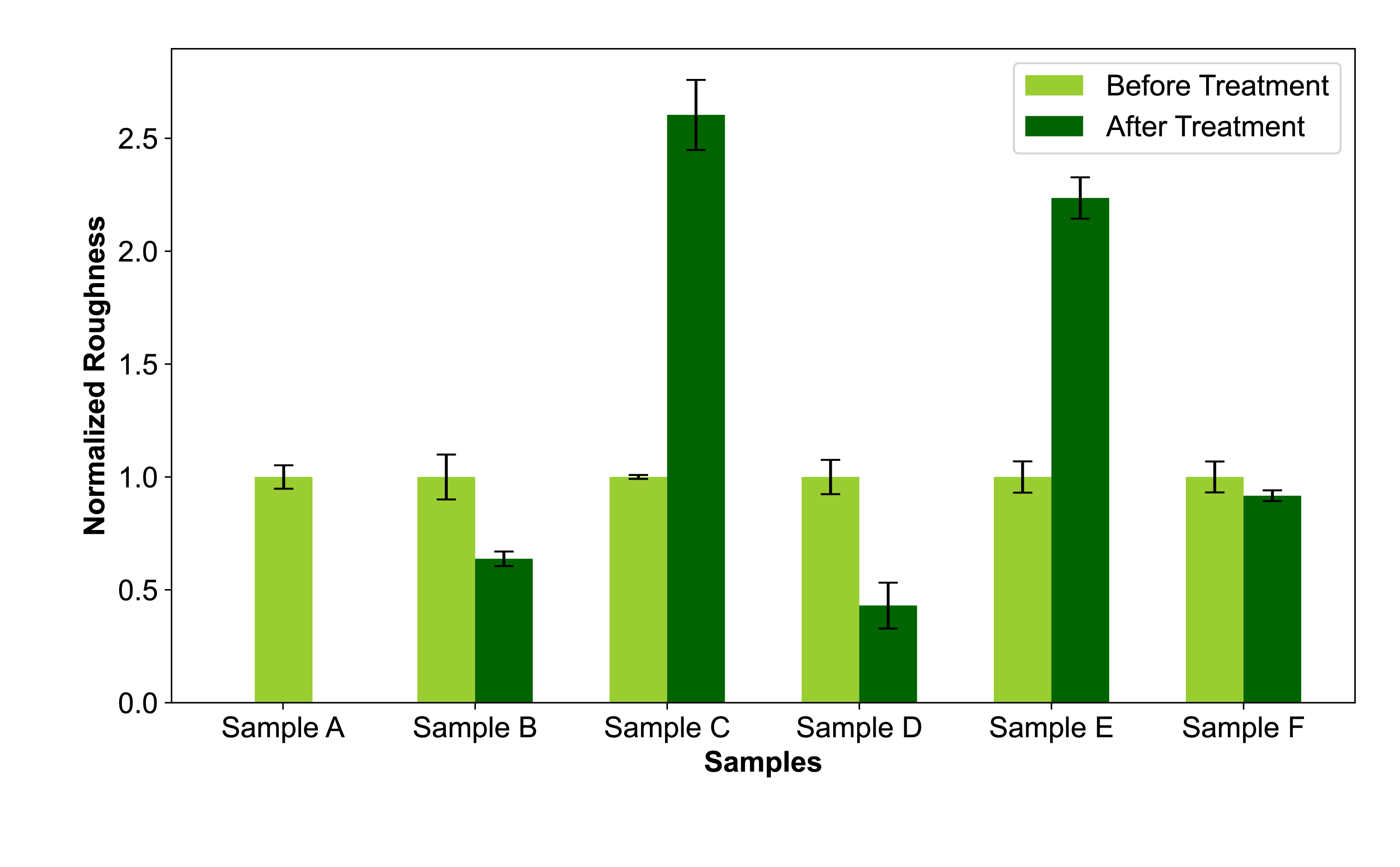}
    \caption{\textbf{Summary of Sample Roughness: }A plot of sample roughness for Samples A - F. The normalized roughness before treatment is shown in light green and the normalized roughness after treatment is shown in dark green. The normalized roughness values for each sample and their respective errors, were calculated by dividing by the initial, pre-treatment roughness.}
    \label{fig:RoughnessPlot}
\end{figure}

\clearpage
\newpage
\section{Energy Dispersive X-Ray (EDS) Mapping of Sample A}
\label{sec:EDS-spectrum}

\begin{figure*}[hbt!]
    \centering
    \includegraphics[width=1\linewidth]{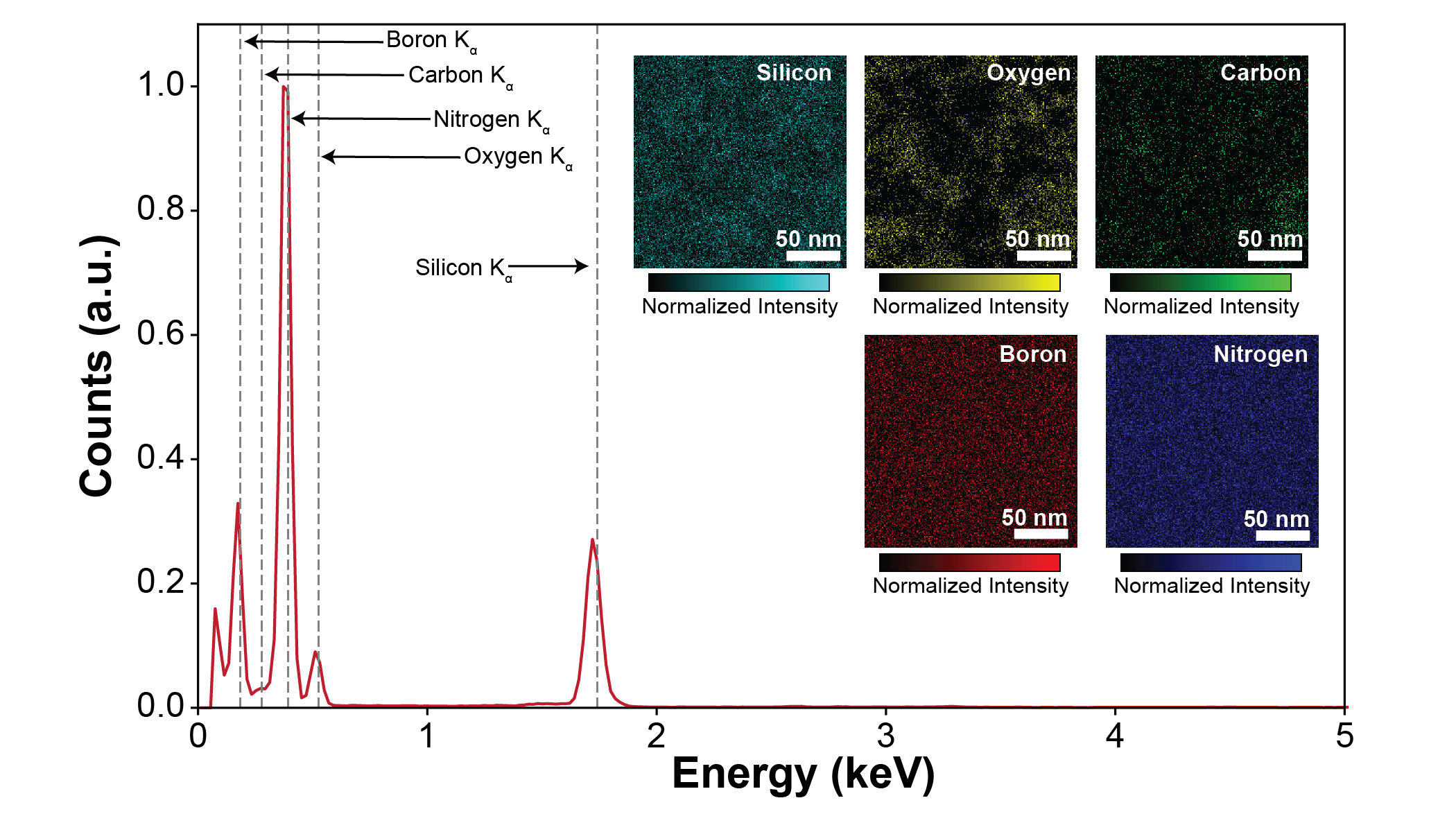}
    \caption{\textbf{EDS Spectra for Sample A: }A plot of the EDS spectra obtained for Sample A. The corresponding elemental maps for silicon, oxygen, carbon, boron and nitrogen are inset.} 
    \label{fig:S1}
\end{figure*}
Figure \ref{fig:S1} shows the EDS spectra corresponding to the elemental maps shown in Figure \MainFigTwo\ in the main text. 
An EDS spectrum image was obtained for Sample A in the JEOL F200, operating in STEM mode with an 80 kV accelerating voltage, and the following parameters: a 0.01 s/px dwell time, a 196 $\times$ 196 px$^2$ image size, and a 20 eV/channel dispersion.
The plotted spectrum shown above was obtained by summing the spectrum collected at each pixel, across all pixels in the spectrum image.
This spectrum image was used to produce the elemental maps for silicon, oxygen, carbon, boron and nitrogen shown in Figure \MainFigTwo\ and inset above. 
These maps were produced in Gatan's GMS software by integrating the intensity of the peak corresponding to each element. 
Each map is normalized with respect to its maximum intensity. 

\newpage
\section{Quantifying Changes in Contamination of Sample A} 
\label{sec:contamination}

\begin{figure*}[ht] 
    \centering
    \includegraphics[width=0.9\linewidth]{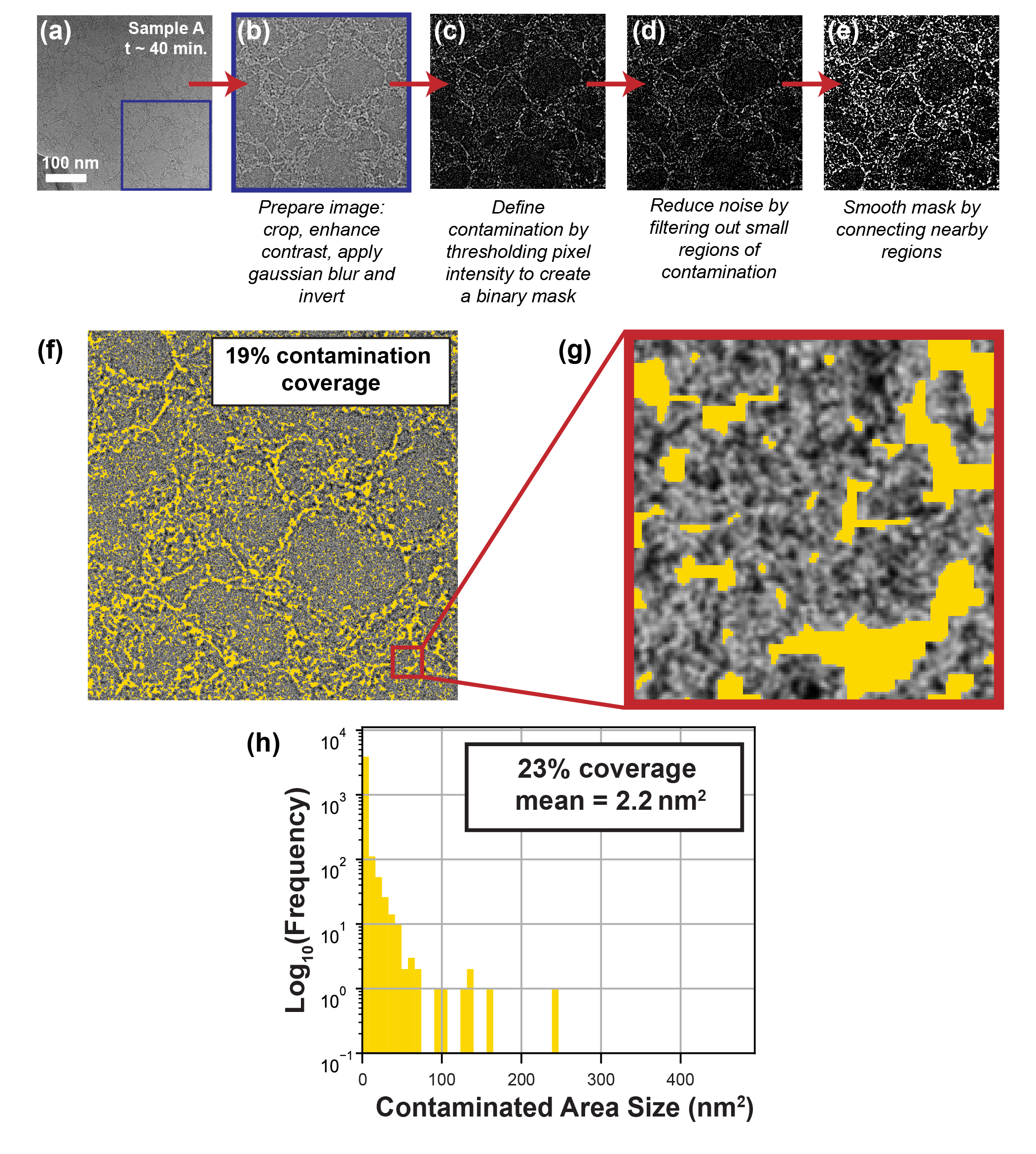}
    \caption{\textbf{Analysis of contamination coverage on Sample A:} (a) Input TEM micrograph of Sample A after $\sim$ 40 minutes of electron irradiation. (b) Processed micrograph after cropping, enhancing contrast with CLAHE, applying a Gaussian blur and inverting the image for contouring. (c) Resulting binary mask after thresholding. (d) The same binary mask after filtering out small, contiguous groups of contaminated (white) pixels. (e) The final binary mask after applying a morphological filter to connect nearby regions of contamination. (f) Overlay of the final binary mask (yellow) on the cropped input micrograph. (g) Zoomed-in view of the region boxed in red in panel (f). (h) Distribution of discrete, contaminated areas shown in (f). }
    \label{fig:S2}
\end{figure*}

TEM micrographs of Sample A show dark, lacey features (see Fig.~\MainFigTwo\ in the main text). 
These features are likely contamination resulting from the hBN transfer process, as indicated by EDS elemental maps.
We identified and quantified these features according to the image processing routine outlined in Figure \ref{fig:S2}. 
First, images are cropped and processed using contrast-limited adaptive histogram equalization (CLAHE) \cite{Pizer1987AdaptiveVariations} to equalize local contrast and remove shadows.
Next, a Gaussian blur is applied to reduce noise and the image is inverted for contouring using OpenCV in Python \cite{Bradski2000OpenCV}, shown in panel (b).
Contaminated areas are subsequently defined by thresholding the image to create a binary mask, shown in panel (c).
The binary mask is improved by filtering out discrete, small contaminated areas below a certain size and using a morphological filter to connect nearby contaminated areas, shown in panel (d).
The final mask can then be overlaid on the cropped input image (panel (f)) and used to compute a percent contamination coverage.
Percent contamination coverage is defined as: 

\[\text{\% contamination coverage} = \frac{\text{number of contaminated (yellow) pixels}}{\text{total number of pixels}} \times 100\% \]
 Based on this analysis, Sample A has a contamination coverage of 19\% after 40 minutes of electron irradiation. 
 We also plotted the distribution of discrete, contaminated areas in Sample A, shown in panel (h) to better visualize changes in contamination over time. 
 The image processing routine described above utilizes several user-defined parameters, which are summarized in the list below. 
\begin{itemize}
    \item Input image size -- 440 $\times$ 440 nm$^2$
    \item Image cropping -- cropped image size, s = 220 $\times$ 220 nm$^2$ -- Images were cropped to only include the lower right quadrant to mitigate the effect of the overall background in the TEM images (see Fig.~2) which made quantification challenging. Cropping also reduced computation time since our processing routine is performed serially, pixel by pixel.
    \item CLAHE -- tile grid size, t = 2 $\times$ 2 nm$^2$ -- Tile grid size defines non-overlapping regions of pixels (tiles) over which CLAHE is applied. A tile grid size of 2 $\times$ 2 nm$^2$ best enhanced contrast without significantly increasing noise in each image.
    \item CLAHE -- clip size, c = 1.1 -- clip size defines the maximum contrast enhancement allowed within a given tile. 
    \item Gaussian blur -- kernel size, k$_{gb}$ = 0.3 $\times$ 0.3 nm$^2$ -- kernel size defines a window that slides across the image, applying the filter to each pixel and its neighbors within the window. 
    \item Gaussian blur -- sigma, $\sigma$ = 0.03 nm -- Sigma defines the width of the Gaussian distribution. 
    \item Thresholding -- threshold intensity, i = 175 -- Pixel intensity ranges from 0 (black) to 255 (white). Pixels with intensity greater than or equal to i become white, and pixels below i become black.
    \item Filtering minimum areas -- minimum area, $\text{a}_{min}$ = 9 nm$^2$ -- This parameter defines a minimum area size for discrete contaminated regions. Regions with areas below a$_{min}$ are set to 0 intensity. This step reduces noise in the final mask. 
    \item Morphological closing filter -- kernel size, $\text{k}_m$ = 1.0 $\times$ 1.0 nm$^2$ -- Similar to k$_{gb}$, k$_m$ defines a sliding window over which the morphological closing filter operates. As this value increases, contaminated regions farther and farther apart are connected (and vice versa).
\end{itemize}

\begin{figure}[ht] 
    \centering
    \includegraphics[width=0.85\linewidth]{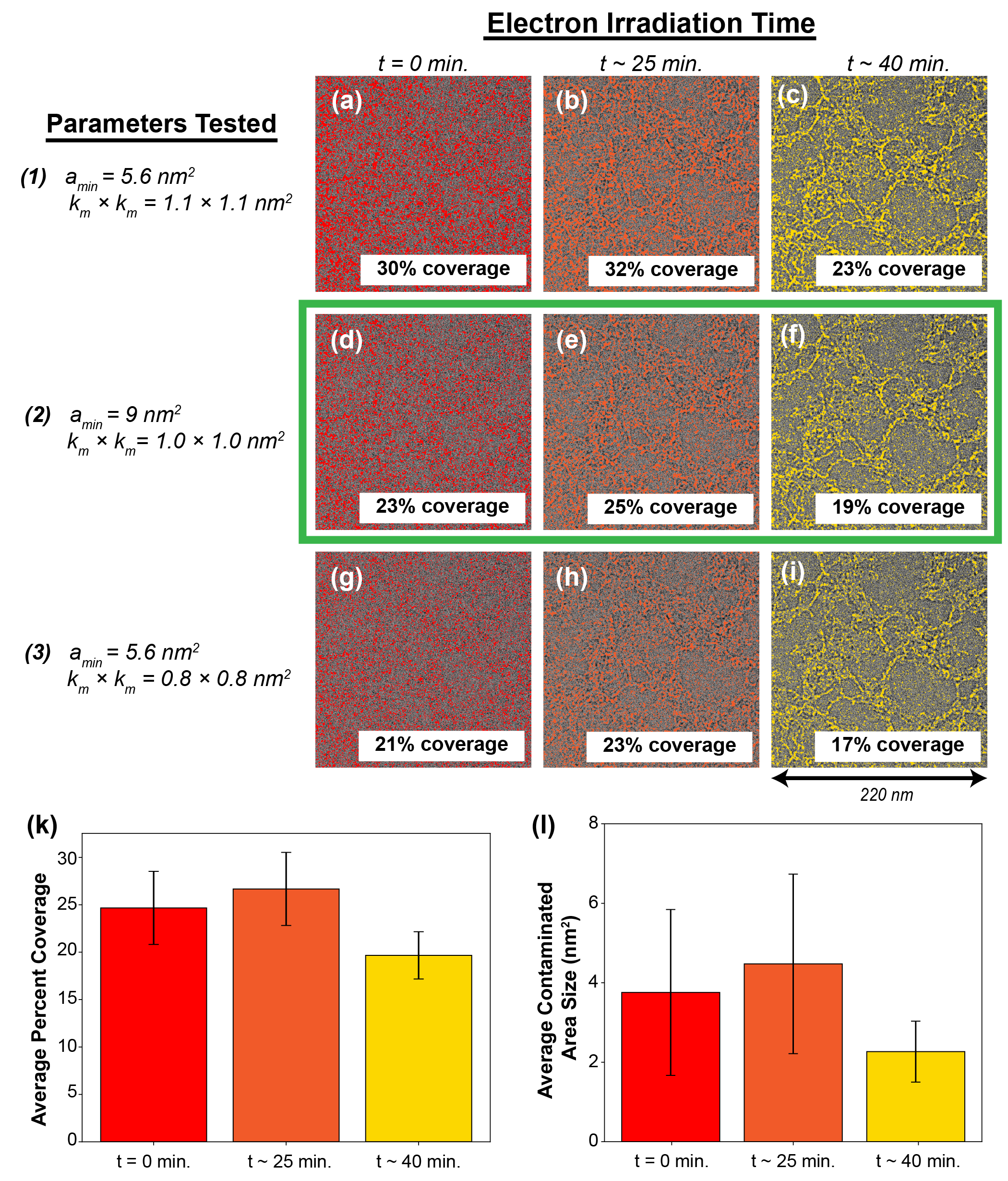}
    \caption{\textbf{Determining Morphological Filter Kernel Size and Minimum Area for Sample A:} (a-c) Contaminated regions identified at 0, 25, and 40 minutes of electron irradiation for the first set of parameters. (d-f) Contaminated regions identified at 0, 25, and 40 minutes of electron irradiation for the second set of parameters. (g-i) Contaminated regions identified at 0, 25, and 40 minutes of electron irradiation for set (3) of parameters. (k) Bar chart showing the average percent coverage, across all three sets of parameters, and the corresponding standard deviations at each time. (l) Bar chart showing the average contaminated area size, across all three sets of parameters, and the corresponding standard deviations at each time.}
    \label{fig:S3}
\end{figure}

As we varied the user-defined parameters across reasonable physical ranges, the results of our analysis only changed by a few percent. The set of parameters that best identified contaminated areas (determined by visual inspection) were ultimately used.
Figure \ref{fig:S3} demonstrates the parameter optimization process for parameters a$_{min}$ and k$_m$. Three sets of a$_{min}$ and k$_m$ are shown in panels (a) through (i). 
The second set of parameters (a$_{min}$ = 9 nm$^2$ and k$_m$ = 1.0 nm), boxed in green, was chosen as the optimal set of parameters. 
The first set of parameters (a$_{min}$ = 5.6 nm$^2$ and k$_m$ = 1.1 nm) tended to overestimate contamination (by including noise) and the third set of parameters (a$_{min}$ = 5.6 nm$^2$ and k$_m$ = 0.8 nm) tended to underestimate contamination (by leaving neighboring contaminated regions disjoint). 
Note that additional combinations of a$_{min}$ and k$_m$, and other parameters not shown here, were tested and compared in a similar fashion.)
Panels (k) and (j) in Figure \ref{fig:S3} show that for different combinations of parameters, percent coverage of contamination and the average size of contaminated regions peak at 25 minutes and are at their lowest at 40 minutes. 

\clearpage
\section{Quantifying Changes in Contamination of Samples B and C} 
\label{sec:contaminationBC}
\begin{figure}[ht] 
    \centering
    \includegraphics[width=1\linewidth]{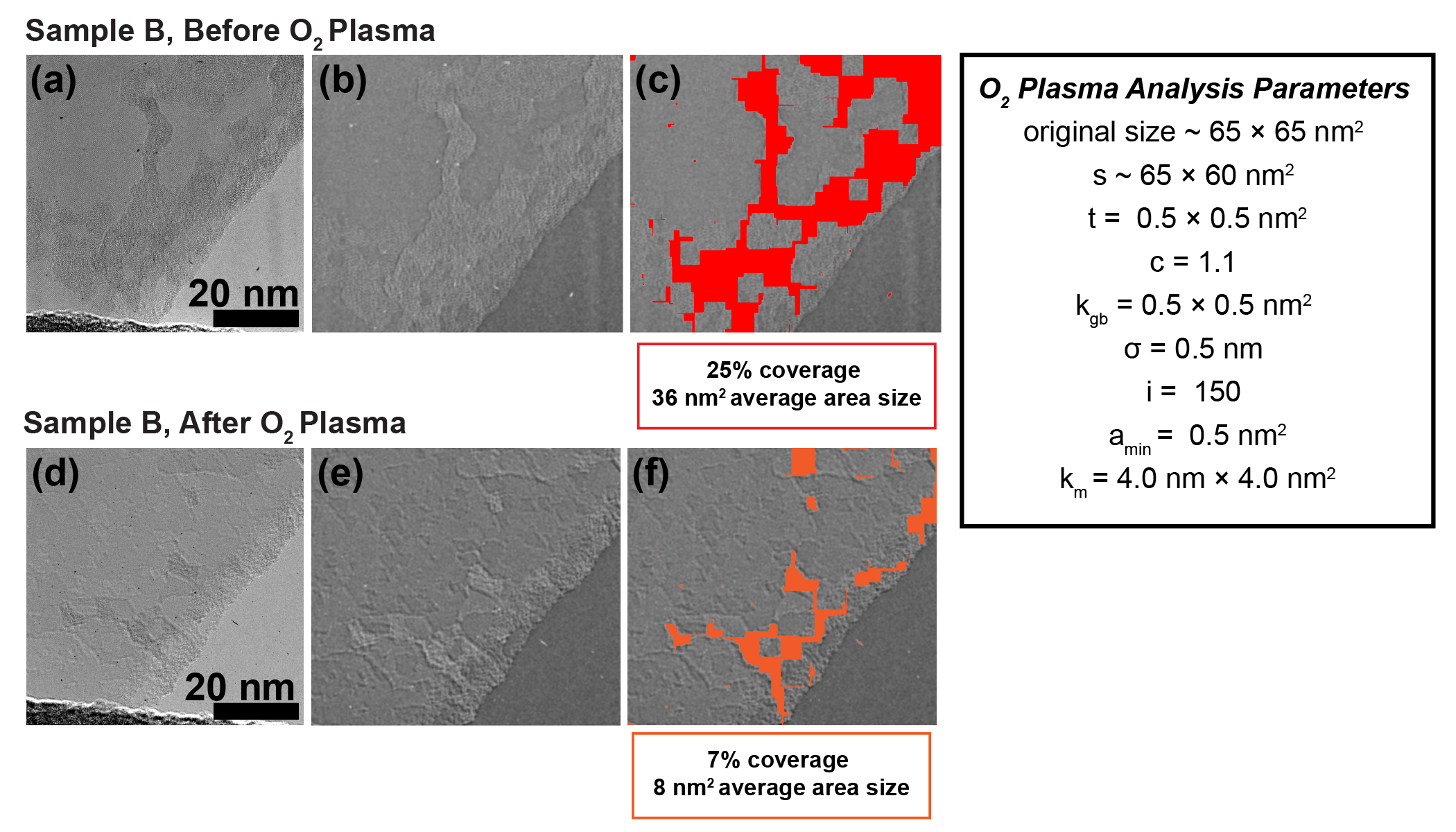}
 \caption{\textbf{Analysis of contamination on Sample B: }(a) Original image of Sample B \textit{before} treatment, shown in Fig.~4 in the main text. (b) Processed version of (a) after cropping, enhancing contrast, applying a Gaussian blur, and inverting pixel intensity. (c) Contaminated regions found in (a), colored in red. (d) Original image of Sample B \textit{after} treatment, shown in Fig.~4 in the main text. (e) Processed version of (d) after cropping, enhancing contrast, applying a Gaussian blur, and inverting pixel intensity. (f) Contaminated regions found in (d), colored in orange.}
 \label{fig:contaminationB}
\end{figure}

\begin{figure}[htb] 
    \centering
    \includegraphics[width=1\linewidth]{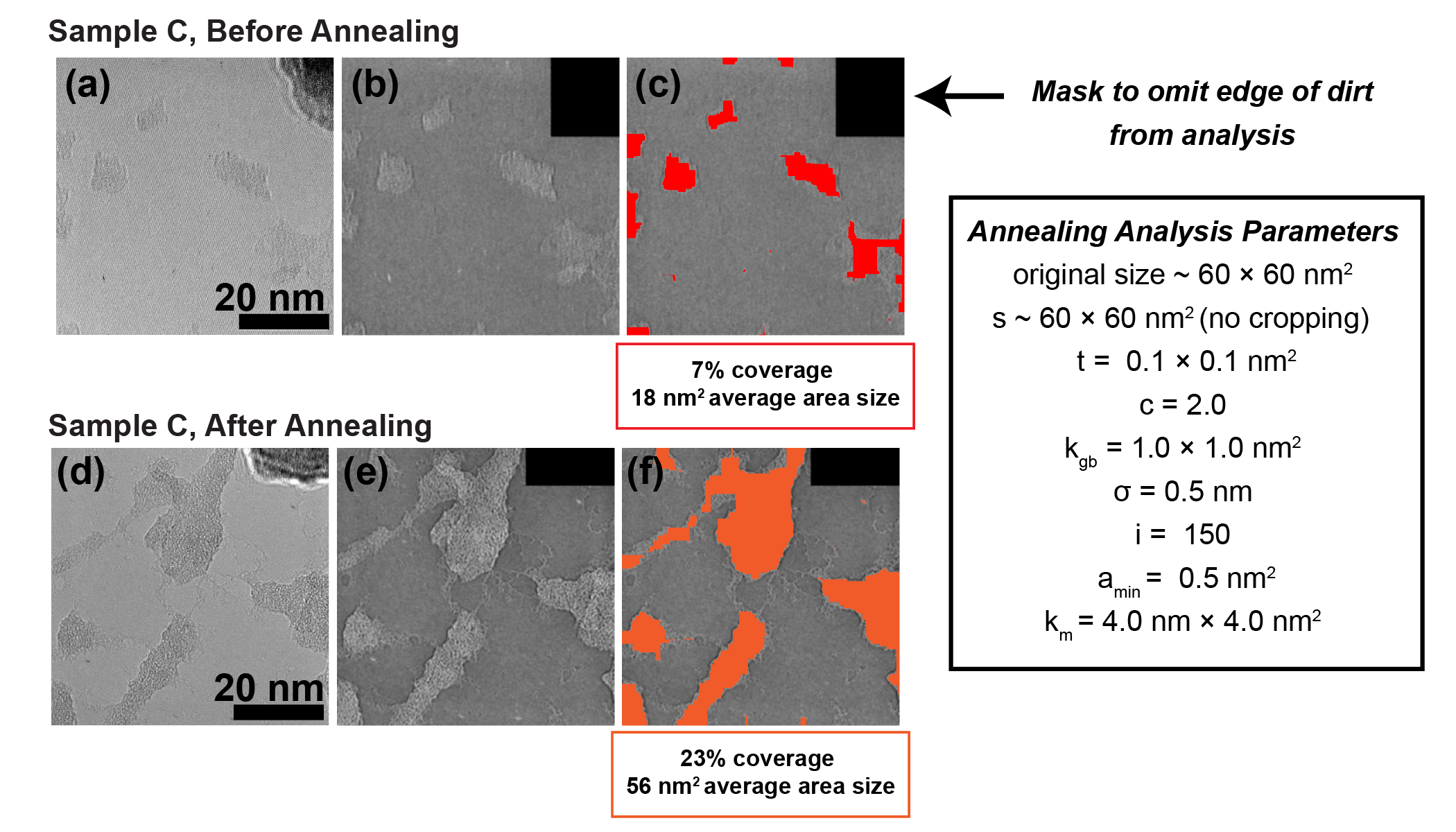}
    \caption{\textbf{Analysis of contamination on Sample C: }(a) Original image of Sample C \textit{before} treatment, shown in Fig.~4 in the main text. (b) Processed version of (a)  after applying a mask (black rectangle in the upper-right corner), enhancing contrast, applying a Gaussian blur, and inverting pixel intensity. (c) Contaminated regions found in (a), colored in red. (d) Original image of Sample C \textit{after} treatment, shown in Fig.~4 in the main text. (e) Processed version of (d)  after applying a mask (black rectangle in the upper-right corner), enhancing contrast, applying a Gaussian blur, and inverting pixel intensity. (f) Contaminated regions found in (d), colored in orange.}
    \label{ContaminationC}
\end{figure}

The same analysis performed on Sample A was repeated for Samples B and C. However, because the images of Samples B and C differ in size and resolution with respect to each other \textit{and} Sample A, new parameters were optimized to detect contamination for these samples. Figures \ref{fig:contaminationB} and \ref{ContaminationC} show the identified contaminated regions, the percent contamination coverage, and parameters used to analyze Samples B and C, respectively. 


\clearpage
\newpage 

\section{Effect of TEM Column Vacuum on Optical Activity of Sample A}
\label{sec:TEM-vacuum}

\begin{figure*}[ht]
    \centering
    \includegraphics[width=.7\linewidth]{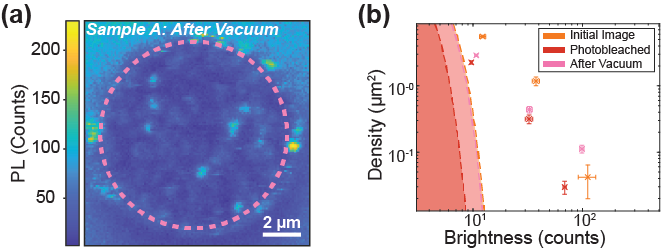}
    \caption{\textbf{PL image after TEM column vacuum:} (a) PL image after being under vacuum in the TEM column. Dashed pink circle indicates suspended region. (b) Emitter family analysis comparing initial and photobleached PL images of Sample A (Fig.~3a,b) to (a).} 
    \label{fig:TEM-vacuum}
\end{figure*}
Sample A was placed in the TEM chamber without exposure to the electron beam for an extended time ($\sim$1.5 hrs) to determine whether the environment of the TEM impacts optical properties.
This test was performed after the photobleaching and before the TEM exposure images shown in the main text (Fig.~\MainFigPlEffects).
Figure \ref{fig:TEM-vacuum} shows the PL image after being in the TEM chamber.
There are no apparent changes to the optical activity in this image compared to the initial images (Fig.~\MainFigPlEffects a,b).
Emitter family analysis confirms there are no significant changes due to the TEM environment and vacuum (Fig.~\ref{fig:TEM-vacuum}b).
Interestingly, the background level increases to be about the same as the initial PL image of Sample A.
The emitter families seem to generally fall in between the families of the initial and photobleached images.

\section{TEM and Treatment Effects on Optical Activity}
\label{sec:SI-PL-effects}

\begin{figure*}[ht]
    \centering
    \includegraphics[width=0.9\linewidth]{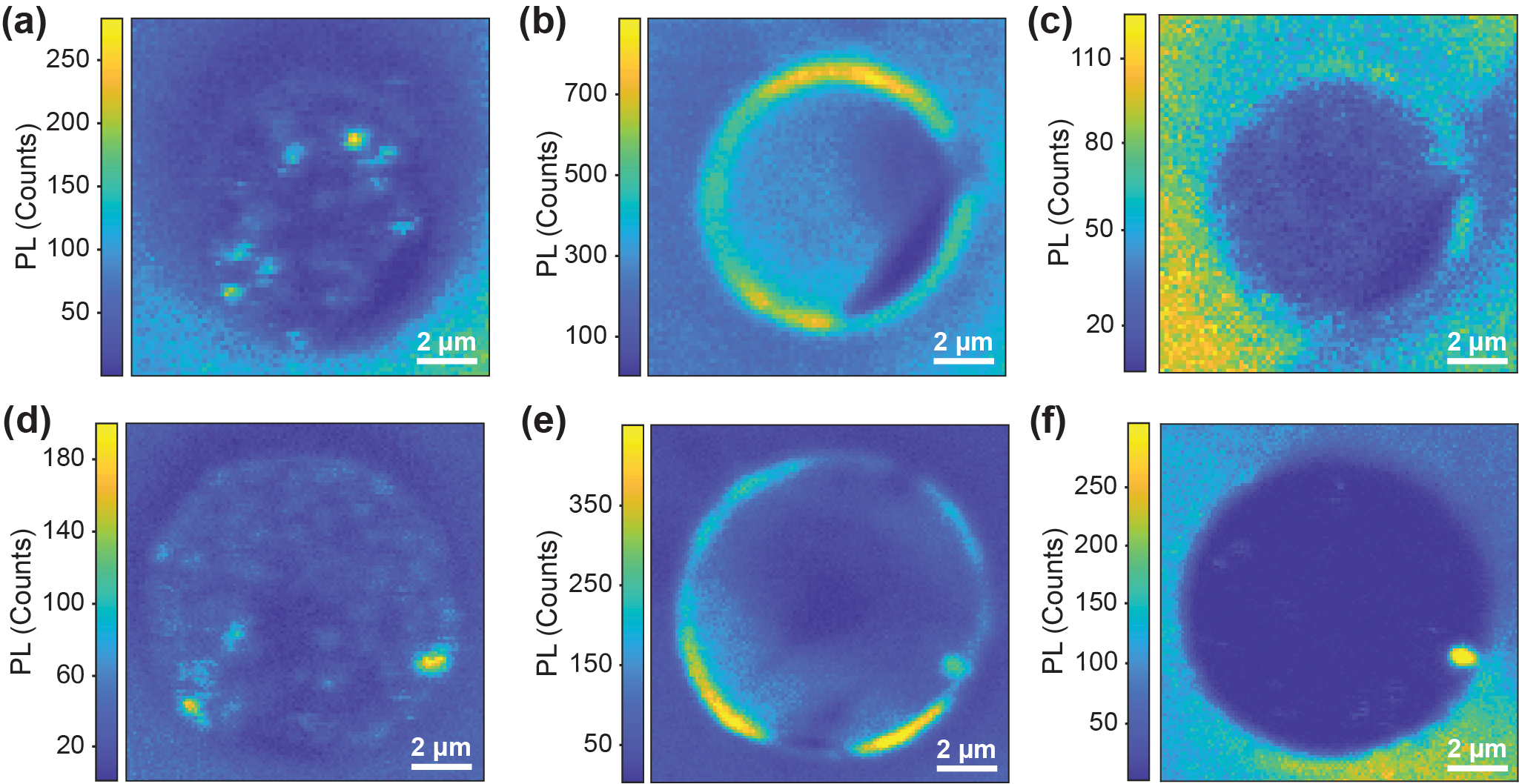}
    \caption{\textbf{PL Measurements of Samples B \& C:} (a) Initial PL image for Sample B. (b) PL image of Sample B after TEM measurement. (c) PL image of Sample B after oxygen plasma treatment. (d) Initial PL image for Sample C. (d) PL image of Sample C after TEM measurement. (e) PL image of Sample C after annealing.} 
    \label{fig:PL-BC}
\end{figure*}
PL images were taken of Samples B and C before and after the TEM imaging and treatments shown in Figure 4 of the main text. 
The initial PL images show the typical localized emission (Fig.~\ref{fig:PL-BC}a,d).
The optical activity changes significantly after exposure to the electron beam (Fig.~\ref{fig:PL-BC}b,e).
The overall brightness greatly increases, and no localized emitters are visible.
The edge of the FIB hole also brightens considerably.
Notably, none of these effects occur when a sample is placed in the TEM column under vacuum without electron beam exposure (see Fig.~\ref{fig:TEM-vacuum}).
The brightening must, therefore, be related to electron beam.

Following the measurements shown in Figures~\ref{fig:PL-BC}b,e, Samples B and C were treated by oxygen plasma and annealing, respectively.
The measurements shown in Figures~\ref{fig:PL-BC}c,f were taken after treatment but before the "after" TEM measurements shown in Figure \MainFigFour.
After oxygen plasma treatment, the background brightness of Sample B returns to the same level as the initial image, but no localized emission is visible (Fig.~\ref{fig:PL-BC}c).
Though the treatment performed was the exact same as that of Sample D (Fig.~\MainFigPLTreatments a-c), the result is quite different.
While Sample D brightens significantly after oxygen plasma treatment, Sample B shows a similar overall brightness to the initial image.
Both Samples B and D show only diffuse emission after the treatment.
After annealing, emission from the suspended region of Sample C is negligible (Fig.~\ref{fig:PL-BC}f).
This result is similar to that of Sample E (Fig.~\MainFigPLTreatments d-f).
The one bright spot in Figure \ref{fig:PL-BC}f is debris on the sample that is visible throughout the PL images (Fig.~\ref{fig:PL-BC}d,e) and the TEM and white light images (Fig. 4.2).

\begin{figure}[ht]
    \centering    \includegraphics[width=0.9\linewidth]{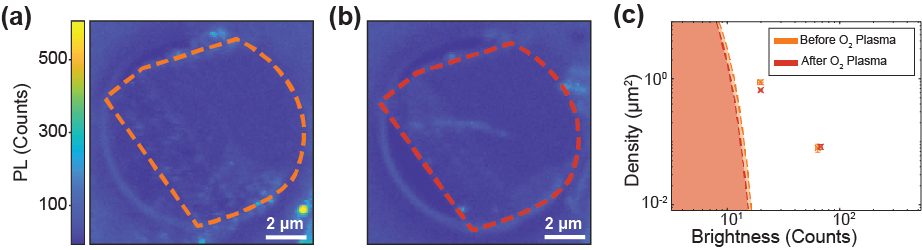}
    \caption{\textbf{PL measurements of alternate oxygen treatment recipe:} (a) Initial PL image of Sample F. (b) PL image of Sample F after oxygen plasma treatment. (c) Emitter family analysis from (a) and (b).}
    \label{fig:alt-oxygen}
\end{figure}
Sample F was exposed to oxygen plasma of 50 W at rate of 25 sccm for 10 mins instead of 50 sccm for 5 mins.
The initial PL image shows that emission is localized and mostly occurs along extended defects (Fig.~\ref{fig:alt-oxygen}a).
Much of this emission is no longer visible after the oxygen plasma treatment, but a new extended defect appears with bright emission (Fig.~\ref{fig:alt-oxygen}b).
Emitter family analysis shows that there is little change in overall optical activity (Fig.~\ref{fig:alt-oxygen}c).
This result is in contrast to the oxygen treatment used on Sample D (Fig. \MainFigPLTreatments a-c).
While the 5 minute treatment results in brightened, diffuse emission, the 10 minute treatment does not result in significant change.
One explanation is that this alternate recipe does not damage the material as much because of the lower flow rate.
This notion is further demonstrated by the change in sample thickness due to the treatment.
Sample B (50 sccm, 5 min) thinned by $\sim$6 nm, while Sample F (25 sccm, 10 min) thinned by $\sim$3 nm.
Another factor to consider is that the barrel asher heats over time, so it is likely that temperature contributes to the difference in result between the two recipes.
This suggests that there may be an optimal set of irradiation parameters which may clean organic residue from flakes without etching the underlying hBN.

\bibliography{SIref}